\documentclass[useAMS,usenatbib,usegraphicx,iop]{emulateapj}

\usepackage{verbatim}
\usepackage{xspace}
\usepackage{rotating}
\usepackage{array}
\usepackage{aas_macros}
\usepackage{url}
\usepackage{amsmath}

\def\halpha{H$\alpha$\xspace}
\def\hbeta{H$\beta$\xspace}
\def\micron{$\mu$m\xspace}
\def\hii{H$\,${\sc ii}\xspace}
\def\hi{H$\,$\textsc{i}\xspace}
\def\rtf{$R_{25}$\xspace}
\def\hiiphot{\textsc{hiiphot}\xspace}
\def\Msun{M$_{\sun}$\xspace}
\def\msun{M$_{\sun}$\xspace}
\def\tauoldv{$\tau_{\mathrm{old,V}}$\xspace}
\def\tauold{$\tau_{\mathrm{old}}$\xspace}
\def\tauyoungv{$\tau_{\mathrm{young,V}}$\xspace}

\def\ieight{$I_{\mathrm{\nu}}(8\mathrm{\mu m})$\xspace}
\def\fha{$F(\mathrm{H\alpha})$\xspace}
\def\feight{$F(\mathrm{8\mu m})$\xspace}

\def\tauscr{$\tau^{\mathrm{scr}}_{\mathrm{V}}$\xspace}
\def\taumix{$\tau^{\mathrm{mix}}_{\mathrm{V}}$\xspace}

\shorttitle{Quantifying non-star formation associated 8\micron dust emission in NGC~628}

\begin{document}

\title{Quantifying non-star formation associated 8\micron dust emission in NGC~628}

\author{Alison F. Crocker\altaffilmark{1\dag},
Daniela Calzetti\altaffilmark{1},
David A. Thilker\altaffilmark{2},
Gonzalo Aniano\altaffilmark{3}, 
Bruce T. Draine\altaffilmark{3},
Leslie K. Hunt\altaffilmark{4},
Robert C. Kennicutt\altaffilmark{5},
Karin Sandstrom\altaffilmark{6},
J. D. T. Smith\altaffilmark{7}}

\altaffiltext{1}{Department of Astrophysics, University of Massachusetts, 710 North Pleasant Street, Amherst, MA, 01003 \email{alison.crocker@utoledo.edu}}
\altaffiltext{2}{Center for Astrophysical Sciences, The Johns Hopkins University, Baltimore, MD 21218, USA}
\altaffiltext{3}{Department of Astrophysical Sciences, Princeton University, Princeton, NJ 08544 USA}
\altaffiltext{4}{INAF-Osservatorio Astrofisico di Arcetri, Largo E. Fermi 5, 50125 Firenze, Italy}
\altaffiltext{5}{Institute of Astronomy, University of Cambridge, Madingley Road, Cambridge, CB3 0HA, UK}
\altaffiltext{6}{Max-Planck-Institut fur Astronomie, Konigstuhl 17, D-69117 Heidelberg, Germany}
\altaffiltext{7}{Department of Physics and Astronomy, University of Toledo, Toledo, OH 43606, USA}
\altaffiltext{\dag}{alison.crocker@utoledo.edu}

\begin{abstract}   

Combining \halpha and IRAC images of the nearby spiral galaxy NGC~628, we find that between 30-43\% of its 8\micron dust emission is not related to recent star formation. Contributions from dust heated by young stars are separated by identifying \hii regions in the \halpha map and using these areas as a mask to determine the 8\micron dust emission that must be due to heating by older stars. Corrections are made for sub-detection-threshold \hii regions, photons escaping from \hii regions and for young stars not directly associated to \hii regions (i.e. 10-100~Myr old stars). A simple model confirms this amount of 8\micron emission can be expected given dust and PAH absorption cross-sections, a realistic star-formation history, and the observed optical extinction values. A Fourier power spectrum analysis indicates that the 8\micron dust emission is more diffuse than the \halpha emission (and similar to observed \hi), supporting our analysis that much of the 8\micron-emitting dust is heated by older stars. The 8\micron dust-to-\halpha emission ratio declines with galactocentric radius both within and outside of \hii regions, probably due to a radial increase in disk transparency.  In the course of this work, we have also found that intrinsic diffuse \halpha fractions may be lower than previously thought in galaxies, if the differential extinction between \hii regions and diffuse regions is taken into account. 

\end{abstract}

\section{Introduction}

The 8\micron dust emission is an attractive star formation rate measure for galaxies, as it provides excellent resolution in nearby low-redshift galaxies. It is also accessible at higher redshift as the rest-frame 8\micron shifts into other accessible infrared bandpasses.  Indeed, the mid-infrared dust emission around 8\micron broadly correlates with star formation rate indicators such as H$\alpha$ \citep{roussel01, peeters04, wu05, kennicutt09}, Pa$\alpha$ \citep{calzetti05, calzetti07}, far infrared \citep{peeters04, dale05} and radio \citep{vogler05, wu05}. However, \citet{calzetti05} draw attention to the non-linearity of the 8\micron dust emission with respect to the number of ionizing photons from \hii regions, raising some concerns about its use as a star formation indicator. Furthermore, \citet{calzetti05} also note that the 8\micron emission is more diffuse than either recombination line emission from ionized gas (\halpha, Paschen-$\alpha$) or 24\micron emission tracing the hot dust localized to star-forming regions. The goal of this paper is to quantify the galaxy-wide contribution to 8\micron dust emission from non-star forming sources as a guide for using such emission as a star formation tracer in both local and high-redshift galaxies.

8\micron dust emission is made up of two main components, aromatic band emission and continuum emission from warm dust grains. 
In general, the aromatic bands lie between 3-17\micron, with some of the strongest bands at 3.3, 6.2, 7.7, 8.6, 11.2, and  12.7 \micron. Each of these bands has been identified with vibrational modes of C-C and C-H bonds of atoms in aromatic rings \citep[see][for a recent review]{tielens08}. Polycyclic aromatic hydrocarbons (PAHs) are now the widely-accepted carriers of the mid-IR band features (\citealp[originally suggested by][]{leger84} and \citealp{allamandola85}), although other hydrocarbon compounds may contribute as well. For simplicity, we will attribute the emission features to PAHs for the remainder of the paper. Because PAH emission is very bright in photo-dissociated regions (PDRs) surrounding sites of active star formation, it was initially surmised that PAH emission might be a very good star formation rate indicator, at least on global scales \citep{roussel01}.

The second component to 8\micron dust emission is continuum emission. Large dust grains may emit at 8\micron if they are heated by a very intense radiation field (significant when the radiation field density is $>10^{4}$ times local). But much more important for the 8\micron emission is the contribution from very small (PAH-sized; $\approx 6$\AA) grains that are stochastically heated to high effective temperatures. The importance of single-photon heating was made clear by Infra Red Astronomical Satellite (IRAS) observations of diffuse regions  which show much more emission at 12 and 25\micron than expected for classically-sized grains, given the low radiation field intensity \citep{boulanger85, boulanger88}. To fit these and other observations, both PAHs and very small grains are now included in dust grain models \citep[e.g.][]{desert90,li01}.

The PAH band emission is now known to have a cirrus component. Band emission associated with the diffuse interstellar medium (ISM) in the Galaxy was first inferred from photometric measurements from the IRAS \citep{boulanger85} and AROME \citep{giard94} and is easily visible in the Spitzer IRAC 8\micron images from the GLIMPSE survey \citep{benjamin03}. Spectroscopic confirmation that the photometric excess (over the expected dust emission) is indeed due to the aromatic emission bands came from the Infra Red Telescope in Space \citep[IRTS;][]{tanaka96} and the Infrared Space Observatory \citep[ISO;][]{mattila96, sakon04}. Using mid-IR spectra of interarm regions, \citet{vogler05} and \citet{sakon07} identify the aromatic bands in the diffuse ISM in the spiral galaxies M83 and NGC~6946, respectively, using ISO and AKARI. Interestingly, these spectroscopic data show distinct variations of the PAH bands (for example, relatively stronger in the 11.3\micron band) in the diffuse regions, when compared to actively star-forming regions \citep{vogler05, sakon04, sakon07}. This band variation signals a difference in the PAH molecule population and/or ionization.

Using Spitzer IRS spectroscopy on the SINGS sample of local star-forming galaxies, \citet{smith07} separate the PAH bands from the rest of the 8\micron dust emission, establishing that typically 80\% of the 8\micron dust emission is from the PAH bands (their Figure~12, bottom panel), with presumably the remainder due to continuum emission from warm, stochastically-heated grains. However, scatter from galaxy to galaxy clearly exists and low-metallicity galaxies in particular tend to have low ratios of PAH band emission to warm dust continuum emission \citep{engelbracht05,cannon06}.

In this paper, we set out to measure the fraction of 8\micron dust emission that is not due to recent star formation. While the line dividing  `recent' star formation from older generations of stars is somewhat arbitrary, here we will count populations of stars under 100~Myr. The goal is to provide a quantitative warning about using the 8\micron dust emission from a galaxy as a direct tracer of star formation. We take the nearby face-on galaxy NGC~628 as a test case and tie this determination to H$\alpha$ emission, which we assume is entirely powered by photoionisation from the youngest stars. In Section~2, we present the H$\alpha$ and Spitzer IRAC data used. Section 3 describes the methods used to first separate 8\micron dust emission related to \hii regions and then emission related to stars under 10~Myr, stars between 10-100~Myr and stars older than 100~Myr. In Section 4, we present and discuss our results, while in Section 5, we test various systematics within our method. Section~6 presents our conclusions. 

\section{Data}
\subsection{NGC~628}

We chose to work on the unbarred Sc galaxy NGC~628 because it is relatively nearby (7.2~Mpc; \citealt{kennicutt11}), large in angular size ($10.5 \times 9.5$ arcmin$^{2}$), virtually face on (i=6$^{\circ}$) and observed in all the bands (\halpha, 8\micron, 3.6\micron) we require. NGC~628 is not classified as an AGN either by optical emission lines \citep{ho97a} or by mid-IR emission lines \citep{goulding09}. Its stellar mass is $3.6\times10^{9}$ \Msun \citep{skibba11} and its current star formation rate (SFR) is 0.68 \Msun yr$^{-1}$\citep{kennicutt11}. Based on spectral energy distribution fitting,  \citet{aniano12} determine that $\sim$11.6\% of NGC~628's infrared luminosity comes from dust heated at high radiation field intensities (U $> 10^{2}$), a typical value among the SINGS sample spirals.

\subsection{\halpha data}

We use \halpha data obtained at the Kitt Peak National Observatory (KPNO) 0.9m telescope in January 1997 and originally presented in \citet{greenawalt98a}. Narrow-band line and broad-band continuum exposures were taken with 16 dithered exposures of 900s and 480s, respectively. The narrow-band filter used is narrow enough (27 \AA\xspace) to exclude contamination from the nearby [N \textsc{ii}] line. 

A \halpha line-only image was obtained after subtracting a scaled broad-band image with the scale factor determined by comparing the broad-band to narrow-band flux ratio of 94 foreground stars. Using the IDL \textsc{MMM} procedure on these flux ratio values, we reject outliers and determine a mean of 0.806 with a standard deviation of 0.036 (84 non-rejected stars). The standard error on the sample mean is thus only 0.0039 (i.e. only 0.5\%); however, this assumes that the foreground stars correctly approximate the narrow-band to broad-band flux ratio of  NGC~628's average stellar population. In light of this possible systematic effect, we will test values changing within the full 1$\sigma$ range of stellar values in Section~5.3 (thus $\pm0.036$ or 4.4\%). 

Flux calibration was based on standard star observations and the conversion to emission measure performed using the formula appropriate for \halpha and a $T=10,000$K plasma: 

\begin{equation}
\frac{EM}{(\mathrm{pc\: cm}^{-6})}=4.858\times10^{-17}\frac{S(\mathrm{H\alpha})}{\mathrm{erg\: s^{-1}\: cm^{-2}\: arcsec^{-2}}}
\end{equation}
The point spread function (PSF) of the final \halpha image has a full-width half-maximum (FWHM) of 1.8 arcsec and the images have an rms noise level of 3.8~pc cm$^{-6}$ with a pixel size of 0.68 arcsec$^{2}$.

\subsection{IRAC data}
The 3.6 and 8 \micron data on NGC~628 were obtained from the SINGS Data Release 5. A tilted plane was fit to both images to subtract the background, after using an iterative algorithm to determine the set of background pixels, see \citet{aniano12} for full details. 

Next, we created a 8\micron dust-only map by subtracting a fraction of the $I_{\mathrm{\nu}}(3.6\;\mathrm{\mu m})$ image (having first convolved it to the 8 \micron resolution) in order to remove stellar emission from the $I_{\mathrm{\nu}}(8\;\mathrm{\mu m})$ image. We subtracted 0.255 times $I_{\mathrm{\nu}}(3.6\;\mathrm{\mu m})$, as in \citet{helou04}, but also check a range of believable subtraction fractions in Section~5.2. We also note that subtracting the 3.6 \micron band will remove slightly more than the stellar emission as the 3.3 \micron PAH band is within the 3.6 \micron IRAC filter. However, the 3.3 \micron feature should be only 0.03-0.09 times the strength of the 8\micron feature \citep{flagey06} which should be around 0.8 of the 8\micron dust emission. So using a factor of 0.255 for the stellar subtraction, we will only cause a 0.6-1.8\% over-subtraction of the dust emission at 8 \micron.

The rms noise level in the non-stellar $\nu$\ieight image (with a pixel size of 0.75 arcsec$^{2}$) is  $1.3 \times10^{-10}$ erg s$^{-1}$ cm$^{-2}$ sr$^{-1}$ and the measured PSF FWHM is 2.82\arcsec \citep{aniano11}. 

\section{Analysis}

In order to separate the 8\micron dust emission linked to recent star formation, we first define \hii regions using the \textsc{hiiphot} code \citep{thilker00, thilker02} on the \halpha data.  \hii regions signal the presence of massive, ionizing stars, which must have ages under 10~Myr. By masking the emission from these regions (see right-hand panel of Fig.~\ref{fig:images}), we calculate the amount of `non-\hii region' emission for both the \halpha and non-stellar \ieight images. 

This method has two caveats.  First, UV photons can escape from \hii regions, heating PAHs outside of their defined boundaries. Secondly, we miss some star-formation by only using \halpha emission. Some low mass young clusters will be under our detection threshold and so included in the non-\hii region emission.  Similarly, a minority of \hii regions maybe very heavily attenuated and thus not detectable in \halpha. Below, we explain how we put limits on these contributions, thereby arriving at an amount of 8\micron dust heated by $>$10~Myr stars.

A further consideration is that stars between 10 and 100~Myr should contribute significantly to the heating of PAHs outside of \hii regions \citep{peeters04}. These stars are related to `recent' star formation, as the term is usually used by the extragalactic community, so we will separately estimate the contribution of these stars to the non-stellar \ieight image, finally arriving at a firm lower limit to how much 8\micron dust emission must be related to purely old stellar populations.

\subsection{Identification of HII regions}
 \textsc{hiiphot} was run on the optical data with default parameters, including a signal-to-noise (S/N) cutoff of 5. Different emission measure (EM; units of pc cm$^{-6}$) slope cutoff values of 1, 1.5, 2, and 4 cm$^{-6}$ were specified. These different $\Delta$EM cutoffs control how far \hii regions grow from initially identified \hii region `seeds'; smaller values of $\Delta$EM result in larger \hii regions. A $\Delta$EM cutoff of 1.5 cm$^{-6}$ is the recommendation from \citet{thilker00}, a value motivated by wanting to recover as much flux spatially linked to an ionizing sources as possible. But we also keep maps based on a wider range of $\Delta$EM cutoff in order to investigate how the definition of an \hii region changes our results (see Section~5.6). 

The noise level and resolution of the data affect how \textsc{hiiphot} finds \hii regions and we investigate these systematics in Section~5.4. Because we wish to define \hii regions on a scale relevant to the 8\micron image with its larger (2.8\arcsec) PSF, the minimum resolution we can achieve with this analysis is a physical scale of 98~pc. Thus we convolve the \halpha image to reach this resolution and also resample to a pixel size of 1.32$\arcsec$ = 43.6 pc pixel$^{-1}$. In the convolved \halpha image, the noise is at a level of 3.8 pc cm$^{-6}$ (1$\sigma$). 

\begin{figure}
\begin{center}
\includegraphics[width=8.9cm]{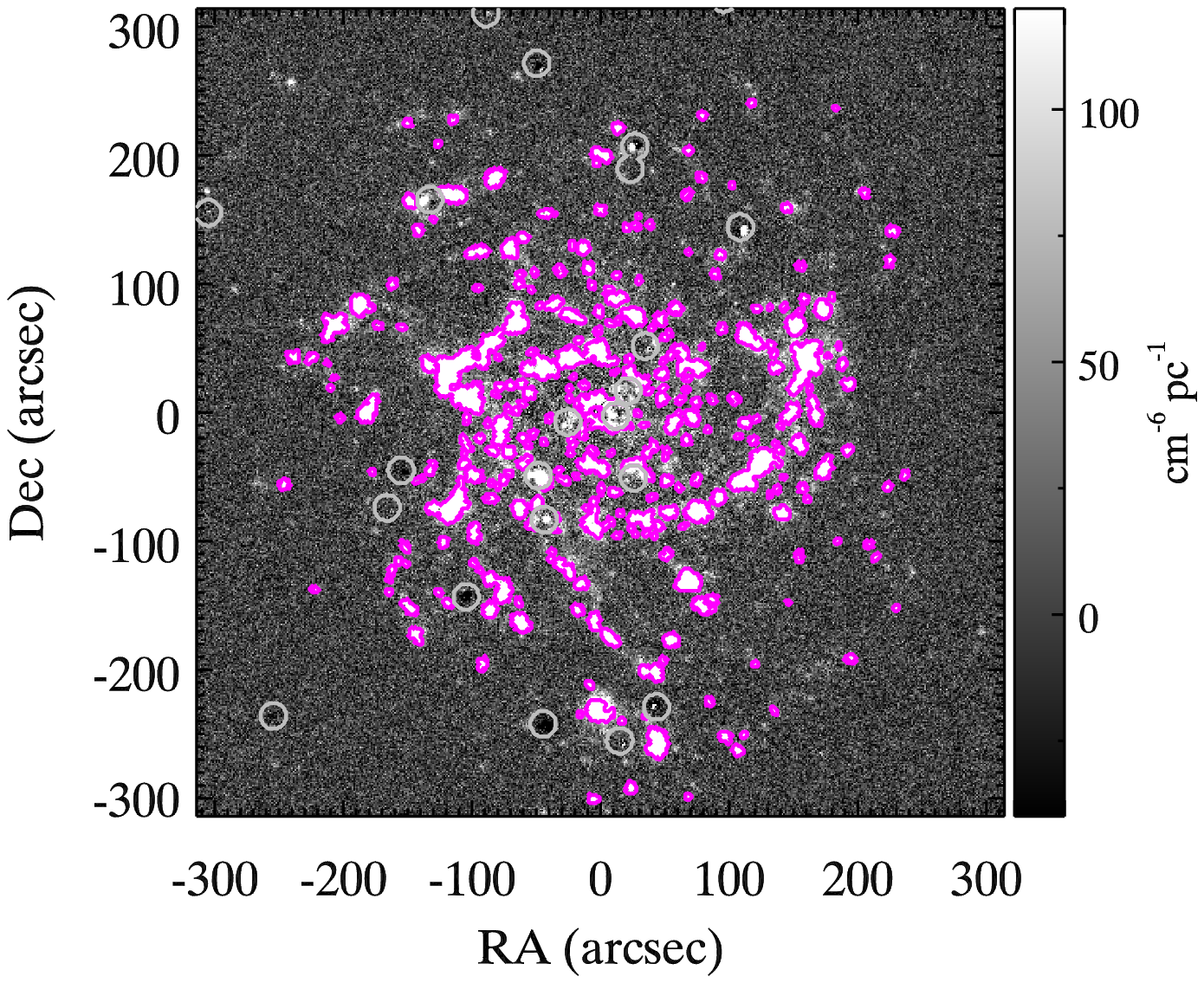}
\includegraphics[width=8.9cm]{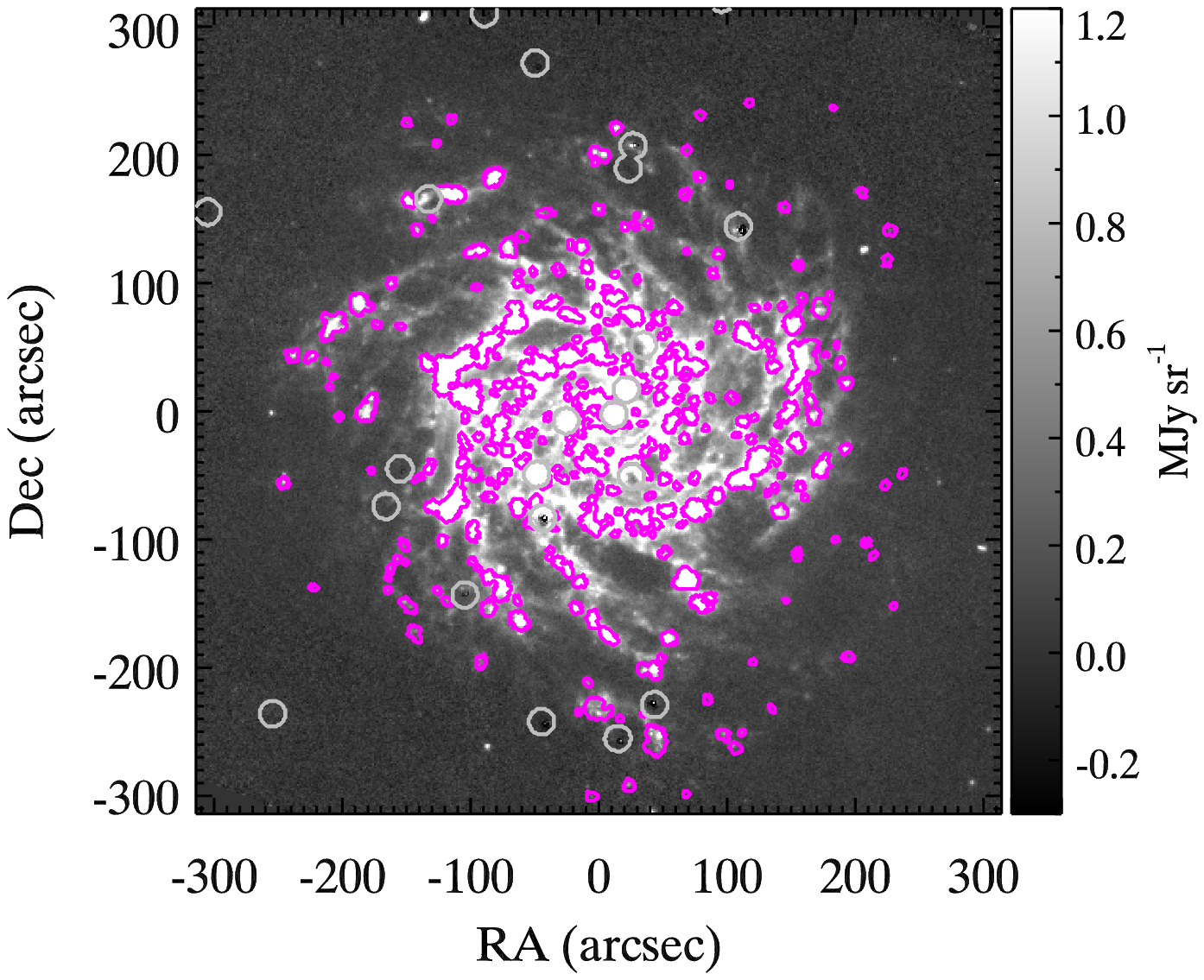}
\caption{\hii regions as determined by \textsc{hiiphot} in magenta contours over the \halpha image (upper) and stellar-continuum subtracted 8 \micron image (lower). Grey circles indicate regions with foreground star contamination, masked out during the analysis.}
\label{fig:images}
\end{center}
\end{figure}

The output of  \textsc{hiiphot}  produces (among other products) an integer mask identifying \hii region pixels (values $>$ 1), non-\hii region pixels (0), and pixels masked from the start (-32768; bright foreground stars). Fig.~\ref{fig:images} shows the outlines of the determined \hii regions (with $\Delta$EM=1.5 cm$^{-6}$ ) on both the \halpha image and over the 8 \micron dust image. 

\subsection{Attenuation correction}

\begin{figure}
\begin{center}
\includegraphics[width=7.8cm]{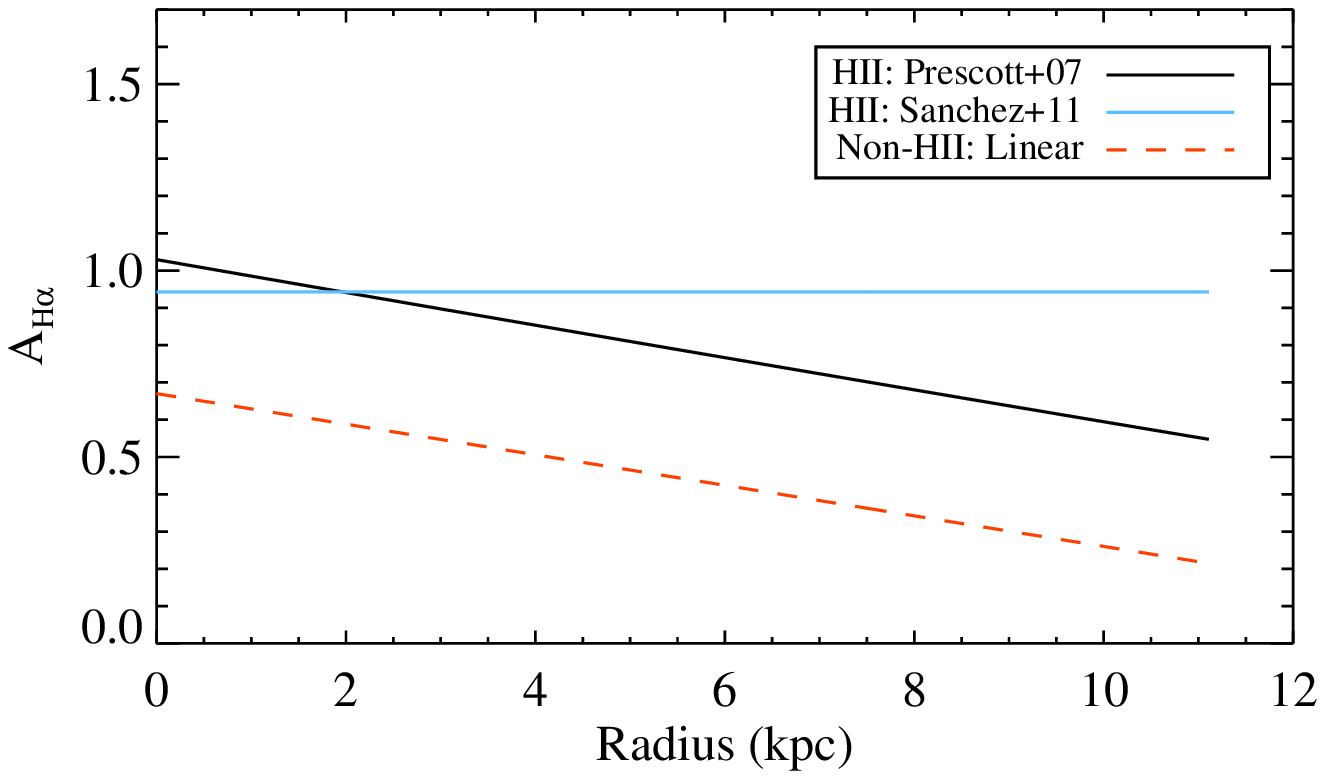}

\caption{The three different extinction corrections used as a function of radius. The solid black line is the \hii region correction from \citet{prescott07} (updated to lower values from \citet{calzetti07} calibration coefficient and the offset in \halpha image calibration), while the solid light blue line shows a constant \hii region correction factor corresponding to 1.24 in A$_{\mathrm V}$ from \citet{sanchez11}. The dashed red line gives the linearly declining extinction correction for the non-\hii regions derived from a dust surface density map with an added contribution from the Milky Way extinction from \citep{schlegel98}. }
\label{fig:Aha_vs_radius}
\end{center}
\end{figure}

The \halpha map must be corrected for \halpha photons lost to dust absorption and scattering. Integral field spectroscopic data of NGC~628 determines V-band attenuations (A$_{V}$) using the Balmer decrement method wherever both \halpha and \hbeta emission lines are detected \citep[mostly \hii regions;][]{sanchez11}. No radial variation in the A$_{V}$ values of these regions is found, with an  average A$_{V}$ of $1.24$ (but large scatter at all radii). Comparing \halpha and 24\micron emission, \citet{prescott07} calculate A$_{V}$ values in 500~pc apertures centered on bright 24 \micron peaks, similarly selecting for \hii regions. Using this method, a radial gradient is found, with A$_{H\alpha}$ declining by -0.53 dex out to \rtf. Figure~\ref{fig:Aha_vs_radius} shows the comparison of the two \hii-region extinctions, together with that for diffuse regions (see below). As both the average gas density and the metallicity decline with radius, we select the radially-declining version of the \hii region extinction correction as the better physically motivated. However, we update the values from \citet{prescott07}, correcting the effect of an incorrectly calibrated \halpha image and using the newer calibration coefficient of \citet{calzetti07} to derive a central A$_{H\alpha}$ of 1.03. 

Evaluating the attenuation in the diffuse regions of galactic disks is even more difficult. Here, we choose to use the dust surface density map from \citet{aniano12} in order to estimate the attenuation in the diffuse regions. Matching the \hii region attenuation, we assume a linear functional form. We find a dust surface density of $2\times10^{5}$ \msun kpc$^{-2}$ in a diffuse region toward the center of the galaxy and $5\times10^{4}$ \msun kpc$^{-2}$ at the edge of the detected region in the dust map, at approximately 9~kpc. Converting to extinction via \citep{aniano12}: 
\begin{equation}
 A_{\mathrm V}[\mathrm{mag}] = 0.67\frac{\Sigma_{M_{\mathrm{dust}}}}{1\times10^{5}\;\mathrm{M}_{\sun}\;\mathrm{kpc}^{-2}},
\end{equation}
we obtain the \halpha attenuation, after additionally converting from V band to \halpha. However, we assume that the emission, on average, originates in the midplane of the galaxy, so we take half of these attenuations, as shown by the red dashed line in Fig.~\ref{fig:Aha_vs_radius}. Additionally, we apply the foreground Milky Way extinction from \citet{schlegel98}.
 
\subsection{Annular photometry}

We choose to analyze the data in large annular regions, in order to have enough signal-to-noise to determine the non-\hii region emission level even in the outer regions of NGC~628. The annuli all have width of $0.1 R_{25} =31.4$\arcsec=1.1~kpc. Using the integer masks produced by \hiiphot, we compute the \hii region and non-\hii region flux in each annular region (hereafter denoted \fha and \feight, where \feight is defined by $\nu F_{\nu}$). Then, to account for diffuse emission contained within the \hii regions but not actually related to the \hii regions, we take the average non-\hii region flux per pixel within each annulus and multiply by the number of \hii region pixels. This value is subtracted off the annular \hii region flux and added to the annular non-\hii region flux for both \halpha and 8\micron data. While the \ieight image is already background subtracted as explained in Section~2.3, a large (width of 50\arcsec=1.8~kpc) outer annulus around the whole galaxy is used for background subtraction of the \halpha image from 100 to 150\arcsec\xspace beyond \rtf. 

Using this method and not yet applying an extinction correction, we determine a non-\hii region \fha fraction of 37\% (commonly called a `diffuse' \halpha fraction). We note that this value is lower (by 7\%) than that reported in \citet{thilker02}, who use identical data and the same code. This difference results from the different accounting of diffuse flux within  the \hii regions. Here, we have subtracted the average non-\hii region surface brightness from within a given \hii region's radial annulus, while \citet{thilker02} use a more sophisticated approach that interpolates the background using information immediately outside each \hii region. As the \halpha surface brightness is generally higher immediately outside of \hii regions, this approach results in higher diffuse levels subtracted for each \hii region. Here, we would only like to subtract the \halpha emission essentially from `in front' of the \hii region, not that linked to it, hence taking the annular average approach.

\subsection{\feight contributions from $<$10~Myr stars}

The non-\hii region 8\micron flux cannot be directly equated with the 8\micron flux produced by stars older than 10~Myr for two reasons. First, photons escape from \hii regions and secondly, low-mass or highly-embedded clusters may not be detectable in the \halpha image.

Unfortunately, integrated UV escape fractions for \hii regions have not received much attention in the literature (line-of-sight attenuations are more the norm), although escape fractions due to ionizing (Lyman continuum) photons have. Such studies generally conclude that leaking ionizing photons power the diffuse \halpha emission, since other potential sources (shocks from mechanical feedback or in-falling gas, photoionization from post-AGB stars and X-ray binaries) appear unable to provide enough ionizing flux \citep[e.g.][]{ferguson96}. Indeed, spectroscopic data confirm that photoionization is dominant, although not the only ionizing source in galaxies \citep{martin97}. How the ionizing radiation permeates so far away from \hii regions remains partly mysterious \citep[e.g.][]{seon09}; but the inhomogeneity of the ISM and the low recombination rate in low density regions probably suffice as explanations \citep{miller93,dove94,zurita02}. \citet{dong11} argue that much of the diffuse \halpha emission may come from regions not in ionization equilibrium, including emission from plasma recombining after the photoionization source is removed.

In any case, escaping Lyman continuum photons are almost certainly accompanied by longer-wavelength UV photons, the photons most effective at exciting PAHs.  While the interception of ionizing versus UV photons is due to different sources (i.e. atomic hydrogen and dust, respectively), if the escape fraction is primarily mediated by physical holes in the cocoon of dust and gas around \hii regions, the escape fractions should be similar. Although the exact ratio between UV and ionizing photon escape fractions is not well constrained, we will assume the fractions are equivalent. Later, we also discuss the results if twice the fraction of UV photons escape compared to the ionizing photon fraction (see Section~5.1). Thus, the non-\hii region \fha also gives a signal about how much UV flux has escaped \hii regions and is available to excite PAHs and heat small dust grains. To estimate the amount of 8\micron emission powered by these escaping UV photons, we take the diffuse \halpha emission and multiply by the \feight/\fha ratio found for the \hii regions.

One caveat to using the non-\hii region \fha to correct for escaping UV photons is that some of the non-\hii region \halpha is scattered off of dust instead of signaling in-situ ionization. Recent studies of high-latitude diffuse regions in the Milky Way estimate that approximately 20\% of the \halpha emission is originally from disk \hii regions and scattered by dust back down towards the disk \citep{witt10,brandt11}. A similar fraction is likely to hold in the non-\hii regions and we apply such a correction here, subtracting 20\% from the non-\hii region \fha before using it to correct for escaping UV photons from \hii regions.

Secondly, we consider the contribution from very young stars ($<$10~Myr) that lie outside of our defined \hii regions. Our minimal \halpha luminosity region detected is $\approx5\times10^{36}$ erg s$^{-1}$. An individual O7 V star produces an \halpha luminosity of $\approx10^{37}$ erg s$^{-1}$, so we can detect regions that only host one or very few ionizing stars. However, some low-mass clusters will not have any ionizing stars and thus not be detected using \hiiphot. We find that \hii regions with low \halpha luminosity have lower \feight/\fha ratios (by about a factor 2) than more luminous \hii regions. This smaller ratio indicates that while these low \halpha luminosity clusters have an ionizing star or two, their IMF is not fully sampled and they have fewer than average bright-UV emitting stars (B-type). Assuming other, equally young, clusters have higher B-star to ionizing star ratios, we would also like to count their contribution to the $<$10~Myr \feight. The twice lower \feight/\fha indicates that we are missing about half of the 8\micron emission that would be measured with a fully sampled IMF. The  total \halpha luminosity of the low-luminosity (but detected) clusters, is only 5\% of the total, so we estimate we need to re-attribute about 2.5\% of the non-\hii region \feight to be associated to $<$10~Myr stars. 

We also likely miss some very highly attenuated \hii regions. However, \citet{prescott07} compared \halpha and 24\micron maps for the SINGS galaxies and determined very few ($<4$\%) such regions exist in normal spiral galaxies. As a rough correction, we subtract 4\% of the \hii region \feight from the non-\hii region \feight and attribute it to \hii region \feight. Combining the sub-ionizing and embedded contributions, the subtraction factor becomes 6.5\%.

We first perform this correction for sub-ionizing young clusters (Equations 3 and 4), then the correction for escaping UV photons (Equation 5 and 6):

\begin{multline}
F(8\mu m)_{\mathrm{nonHII}} \\
= F(8\mu m)_{\mathrm{nonHII, orig}} - 0.065\times F(8\mu m)_{\mathrm{HII, orig}},
\end{multline}

\begin{equation}
F(8\mu m)_{\mathrm{HII}} = 1.065\times F(8\mu m)_{\mathrm{HII, orig}} 
\end{equation}
\begin{multline}
F(8\mu m)_{> 10\mathrm{Myr}} \\
= F(8\mu m)_{\mathrm{nonHII}} - F(H\alpha)_{\mathrm{nonHII}} \frac{F(8\mu m)_{\mathrm{HII}}}{F(H\alpha)_{\mathrm{HII}}},
\end{multline}
\begin{equation}
F(8\mu m)_{< 10\mathrm{Myr}} = F(8\mu m)_{\mathrm{tot}} - F(8\mu m)_{> 10\mathrm{Myr}}.
\end{equation}

\subsection{\feight contributions from 10 to 100~Myr stars}

Stars remain UV bright and thus effective at heating PAHs and small grains up to approximately 100~Myr. Such young stars are still tied to recent star formation, so we would like to estimate their contribution to the 8\micron dust emission.  For both the PAH and dust heating, we estimate the contribution based upon their expected UV flux (912-3000\AA\xspace) of 10-100~Myr stars. Using \citet{bruzual03} models, with a solar metallicity and Chabrier IMF, we find that the expected UV flux of stars between 10-100~Myr is 0.49 times that of the stars younger than 10~Myr. Note that these models are based on an exponentially decaying SFR with a timescale of 5~Gyr \citep[eqn. 6 of][]{noeske07b}. 

Assuming that the UV photons of 10-100~Myr stars are equally effective at producing 8\micron dust emission, then 0.49 times the \feight from $<10$~Myr stars should be produced by 10-100~Myr stars. So, we have:
\begin{equation}
F(8\mu m)_{< 100\mathrm{Myr}} = 1.49 \times F(8\mu m)_{< 10\mathrm{Myr}}.
\end{equation}
However, this almost certainly leads to an overestimate because the 10-100~Myr stars are probably less effective at heating PAHs and dust for several reasons. First, they are generally further away from dust clouds so the number of their photons intercepted by dust is likely lower. Secondly, the older stars' UV emission is not as blue and thus will have a reduced cross-section. Thirdly, some of the 8\micron dust emission from \hii regions is actually a secondary product of the Lyman continuum photons, with PAHs excited by Lyman-$\alpha$, Balmer lines, and 2-photon continuum. As we are most concerned with providing a lower limit to the amount of \feight from $>$100~Myr stars, we keep the overestimate as a safe approach, thus:
\begin{equation}
F(8\mu m)_{\mathrm{oldstars}} > F(8\mu m)_{\mathrm{total}}-1.49 \times F(8\mu m)_{< 10\mathrm{Myr}} .
\end{equation}

\begin{table}[htdp]
\caption{Integrated fluxes}
\begin{center}
\begin{tabular}{lc}
\hline
\hline
Quantity & Flux \\
& (erg s$^{-1}$ cm$^{-2}$) \\
\hline
$F(8\mu m)_{\mathrm{total}}$ & $1.24 \times 10^{-9\phantom{0}}$\\
$F(8\mu m)_{\mathrm{nonHII}}$ & $8.75 \times 10^{-10}$\\
$F(8\mu m)_{> 10\mathrm{Myr}}$ & $7.86 \times 10^{-10}$ \\
$F(8\mu m)_{> 100\mathrm{Myr}}$ & $5.65 \times 10^{-10}$ \\
\hline
$F(\mathrm{H}\alpha)_{\mathrm{total}}$ & $2.65 \times 10^{-11}$\\
$F(\mathrm{H}\alpha)_{\mathrm{nonHII}}$ & $6.35 \times 10^{-12}$\\
\hline

\end{tabular}
\end{center}
\label{default}
\end{table}%

\section{Results}

\subsection{Integrated values}

Excluding star-forming regions identified by \hiiphot, we arrive at a total (out to \rtf) non-HII region dust \feight fraction of 71\% in NGC~628. Then, correcting for other contributions from very young ($<$10~Myr) stars we arrive at a fraction of 62\%. Considering the amount of UV produced by 10-100~Myr stars (see Section~3.5), we estimate a lower limit to the amount of 8\micron dust emission heated by sources other than young stars as 43\%. These are all much higher values than the non-\hii region \fha fraction, determined to be only 24\%, after applying the differential extinction correction. 

Thus the non-star forming component of the 8\micron dust emission is significant compared to the emission directly associated with young stars in NGC~628. Star formation rates computed based on the total 8\micron dust emission thus trace partly the current star formation rate tied to the O and B stars, but also take into account emission that is heated by the integrated population of stars found in the field. This latter contribution may change from galaxy to galaxy and thus pollute the estimates of star formation rates. Future work on additional galaxies will show if this is the case.

\subsection{Radial trends}

\begin{figure}
\begin{center}
\includegraphics[width=9cm]{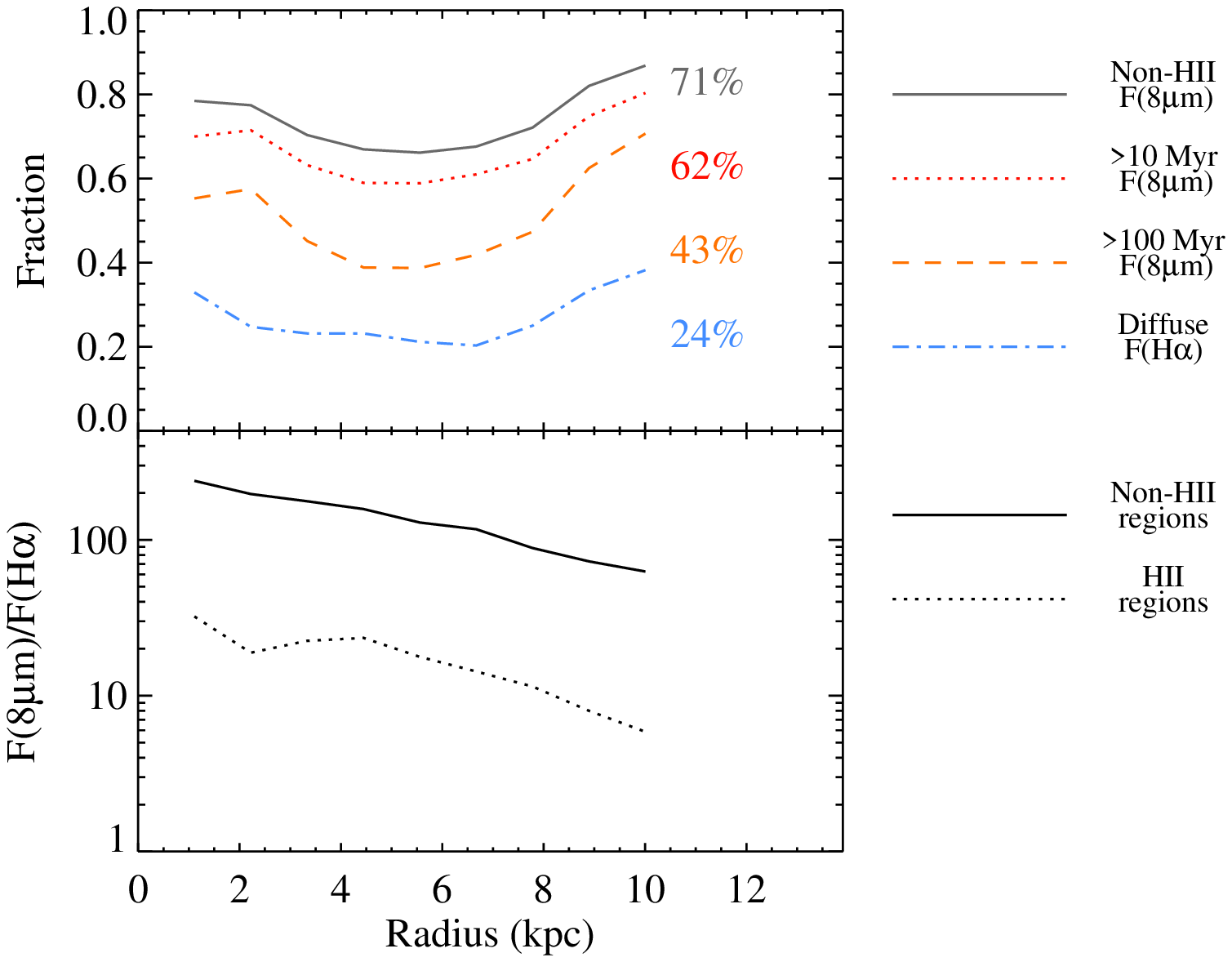}
\caption{\feight fractions and \feight/\fha ratios as a function of radius.
The top panel shows  the non-HII region \feight(gray), the  $>$10~Myr \feight (red), and the lower bound for the $>$100~Myr \feight (orange). The attenuation-corrected non-\hii region \fha fraction is also shown (blue). The total percentages of each quantity are shown on right-hand side of the plot. The bottom panel shows the \feight/\fha ratio as it declines with radius for the non-\hii (solid line) and \hii (dotted line) regions. }
\label{fig:results}
\end{center}
\end{figure}

Fig.~\ref{fig:results} shows the radial trends of the dust \feight fractions contained outside of \hii regions (grey line), heated by stars older than 10~Myr (red dotted line) and heated by stars older than 100~Myr (orange dashed line). The blue dot-dashed line shows the fraction of \halpha emission located outside of \hii regions. 

For all of the \feight fractions, we see a dip with a minimum at 4.4~kpc, the radius at which the brightest \hii regions are found in NGC~628. With the 8\micron dust emission dominated by these bright regions in this annulus, a lower fraction of non-star formation related 8\micron dust emission is expected. After this dip, a gradual increase with radius is seen. This increase is probably due to the relatively greater amount of quiescent gas (and dust) at larger radii. Thus as star forming regions become more sparse in the outer regions of the galaxy, photons from older stars become correspondingly more dominant in producing 8\micron dust emission. 

We also note the decline of the \feight/\fha ratio for both the \hii regions and the non-\hii regions, seen in the bottom panel of Fig.~\ref{fig:results}.  This decline is expected because the gas density declines with radius and thus the disk transparency (assuming a relatively constant dust-to-gas ratio) increases. There are simply fewer PAH molecules to absorb stellar photons, while there are still sufficient hydrogen atoms to absorb $h\nu > 13.6$ eV photons. Potentially, there is also a contribution from the metallicity gradient within the disk as PAHs are observed to be deficient at low metallicities \citep[e.g.][]{madden00,engelbracht05,galliano05}. This observed radial trend highlights the importance of performing this analysis in radial bins.

\subsection{\feight/\fha ratios in individual \hii regions}

\begin{figure}
\begin{center}
\includegraphics[width=8.7cm]{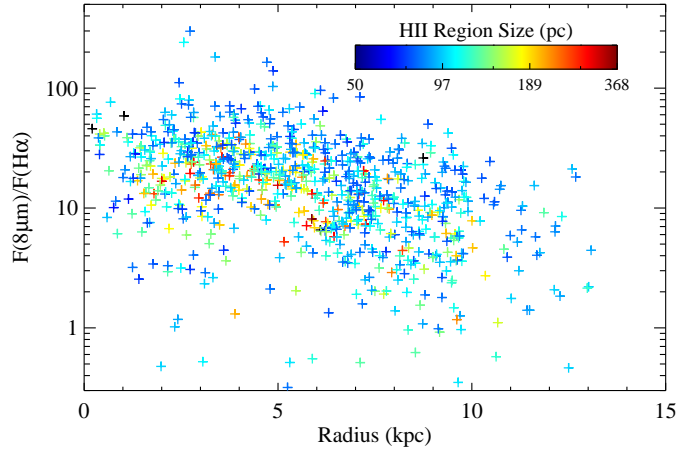}
\caption{\feight/\fha ratio for individual \hii regions as a function of their galactocentric radius. As with the annular average, a clear decline is seen with increasing radius. No relation with HII region size (shown by symbol color) is seen.}
\label{fig:HII_size}
\end{center}
\end{figure}

Fig.~\ref{fig:HII_size} shows the \feight/\fha ratios of individual \hii regions against their galactocentric radius. To make this plot, only HII regions with 8\micron surface brightnesses higher than 10\% above the subtracted background were used (832 of 1055 regions), avoiding those regions whose individual 8\micron flux is not well determined. As previously seen with the annular mean,  the ratio clearly declines with radius, but there is a very large spread at all radii (up to a factor 100). Neither \hii region size (shown as symbol color in Fig.~\ref{fig:HII_size}) nor \halpha luminosity (not shown, but similar appearance) appear to correlate with the \feight/\fha ratio.

\subsection{Fourier power spectrum analysis}
\begin{figure}
\begin{center}
\includegraphics[width=8.7cm]{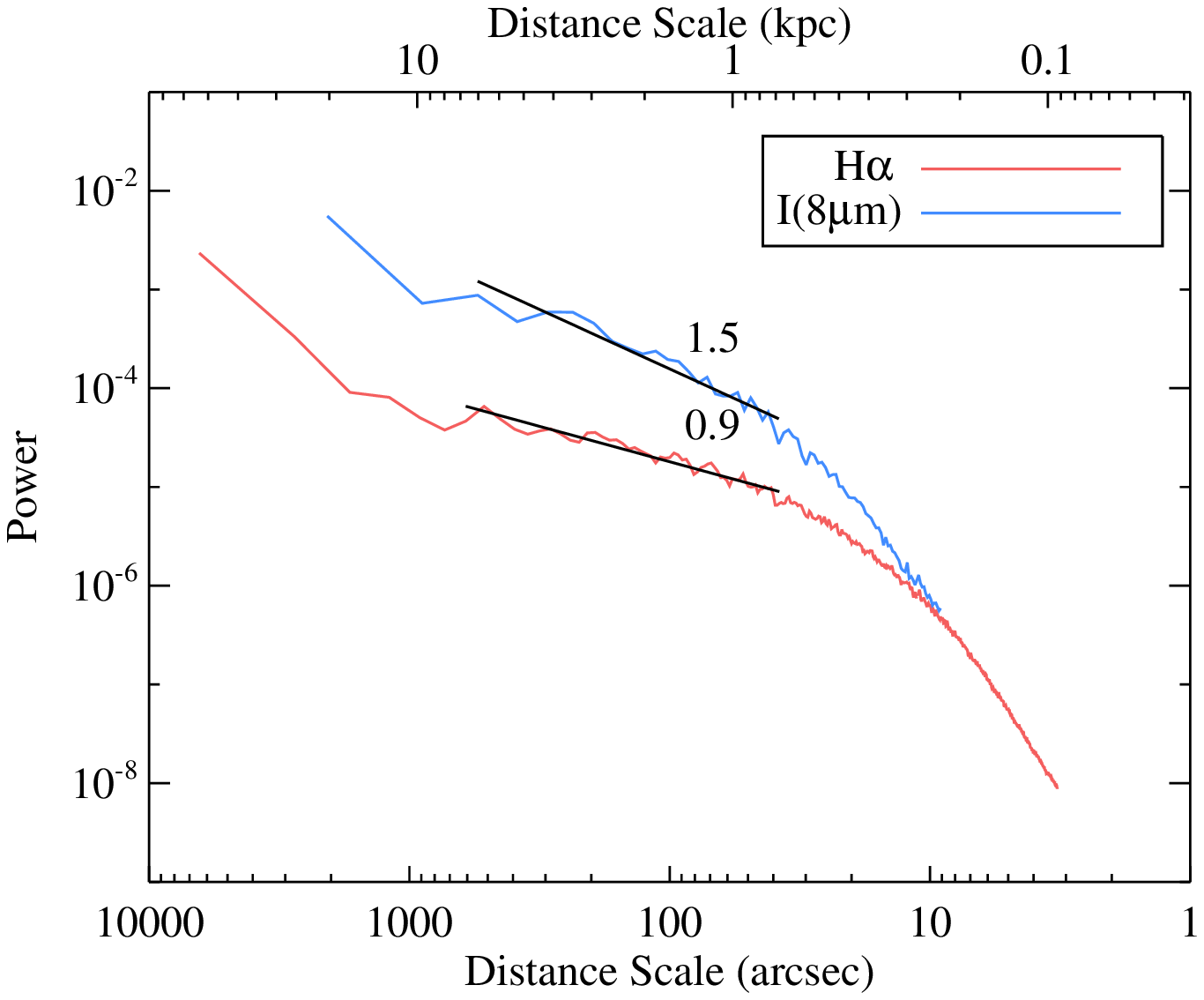}
\caption{Fourier power spectrum of the \halpha (blue) and \ieight dust (red) images. At large scales, the 8\micron dust emission has a clearly steeper slope than the \halpha emission. The steeper slope signals more relative power at large scales, confirming that more of the 8\micron dust emission is spread on larger scales than the \halpha emission. As we are primarily concerned with the slopes, we have not applied any particular normalization to the power spectra.}
\label{fig:powerspec}
\end{center}
\end{figure}

As an independent test, we perform a Fourier power spectrum comparison between the \halpha and \ieight images. We use the native-resolution and original pixel scale images, deprojected to a face-on orientation and ignoring scales below twice the PSF FWHM transformed into the deprojected plane.  We use IDL to compute the 2-dimensional discrete Fourier transform and then take the square of this complex array to arrive at the power spectrum image. We azimuthally average this 2-dimensional power spectrum to obtain the power at each linear scale. This 1-dimensional average power spectrum is shown in Fig.~\ref{fig:powerspec} for both \halpha and \ieight dust images. 

The slope of the \halpha power spectrum is clearly shallower than that of the 8\micron dust emission. We choose to fit the power spectrum slope from 20-200 arcseconds ($\approx$ 0.7-7~kpc) in both images, as both appear approximately power-law in this region. The power-law slopes are $-0.9$ and $-1.5$ for \halpha and \ieight dust images, shown as solid lines in Fig.~\ref{fig:powerspec}, respectively. The value found for the 8\micron dust image is comparable to the $-1.6$ slope found for the \hi emission over similar (0.8-8 kpc) scales by \citet{dutta08}. This similarity signals that there is indeed 8\micron dust emission distributed like the neutral (and not necessarily star-forming) gas.

The $-0.9$ power-law slope found for the \halpha image is somewhat similar to two studies that find slopes of $-0.7$ and $-0.8$ for the large-scale power-law slope for \halpha images of M33 \citet{elmegreen03b, combes12}. However, not all galaxies have such shallow \halpha slopes, in the same paper, \citet{elmegreen03b} report a slope of $-1.5$ for NGC~5055 which is closer to the prediction for a turbulent medium (predicted value of $-1.7$). 

In NGC~628, the relative steepness of the dust \ieight power spectrum compared to that of the \halpha image shows that the 8\micron dust emission is distributed on larger scales than the \halpha emission. This result fits nicely with the larger non-\hii region dust \feight fraction than for the non-\hii region \fha fraction we find in the above analysis. 

\subsection{A simple model}

It has long been suggested that old stars significantly contribute to the FIR emission of galaxies, both from studies of our own Galaxy and external galaxies \citep[e.g.][]{lonsdalepersson87, buat88, trinchieri89, bloemen90, sodroski97, draine07b}. Furthermore, \citet{groves12}  have recently demonstrated that exclusively old stars heat the dust found within the bulge of M31. Here, we demonstrate the plausibility 
 that stars older than 100~Myr produce a significant fraction of PAH emission in NGC~628, while the more detailed calculations appear in the Appendix.
 
 We model the stellar emission of NGC~628 using the evolutionary synthesis code of \citet{bruzual03} and four possible star formation histories (constant, linearly decreasing, exponentially decreasing, peaked at 2~Gyr). The parameters for the linearly decreasing and 2~Gyr-peaked models are based upon \citet{fraternali12} and \citet{zou11}, respectively, while a decay time of 5~Gyr is used for the exponential model. For each stellar age bin, we then calculate the absorption of this starlight by applying the dust absorption cross-section from the \citet{draine07a} Milky Way dust model with a PAH mass fraction of 4.6\%. The scaling of the dust optical depth is tied to the optical attenuation, as determined in Section~3.2. The dust geometry is assumed to be a shell for the very youngest stellar populations, fading into a mixed star and dust geometry for the older regions, although we note that uniformly applying a mixed geometry produced nearly identical results. The fraction of dust absorption due to PAHs at each wavelength is separated by comparing the PAH and total dust cross-sections. This allows us to capture the effect of increased UV cross-section of the PAHs compared to larger dust grains.
 
\begin{figure}
\begin{center}
\includegraphics[width=8.7cm]{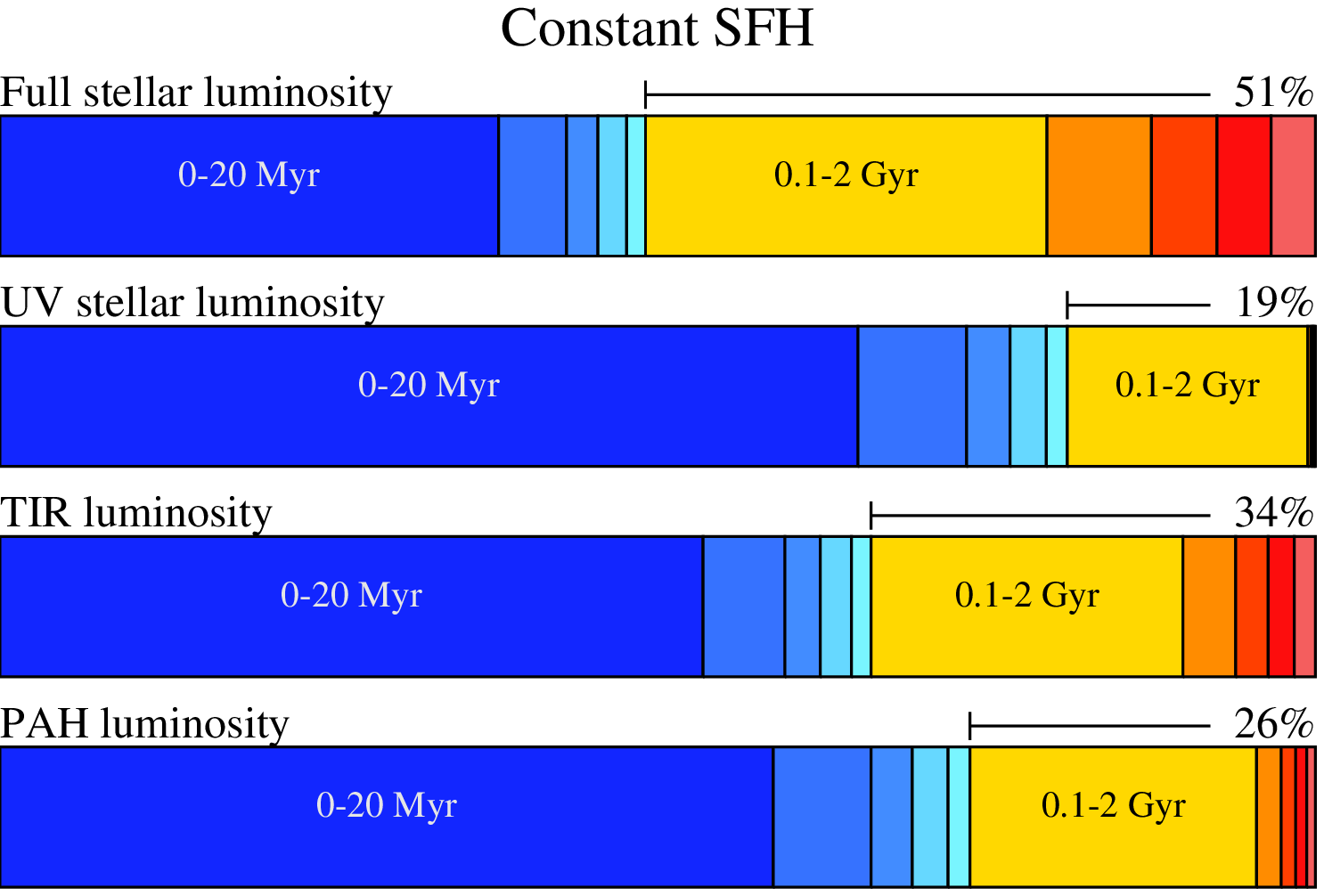}
\includegraphics[width=8.7cm]{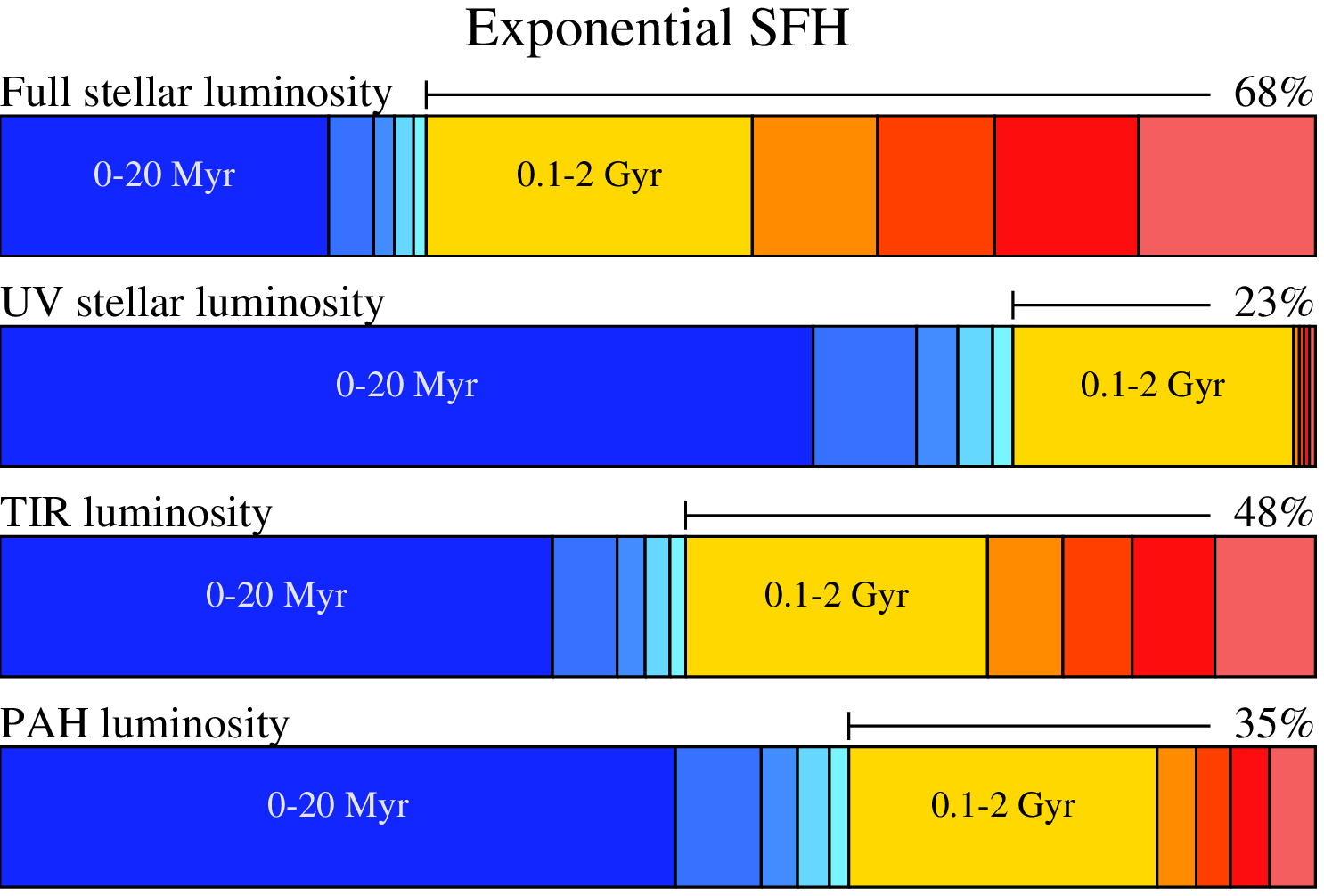}
\caption{The fractional contribution of each age bin to the total stellar, FUV (912-3000\AA\xspace) stellar, TIR and PAH luminosity. The age bins for the young populations ($<$ 100~Myr) are shown in shades of blue and each bin represents 20~Myr. The age bins for the old populations are shown in shades of red and each bin represents 2~Gyr (other than the first, which is 1.9~Gyr). The top four bars represent a constant SFH stellar population while the bottom four bars represent an exponential stellar population with timescale 5~Gyr. }
\label{fig:fractional}
\end{center}
\end{figure}

Figure~\ref{fig:fractional} shows the fractional contributions of different age bins for two different star formation histories (constant and exponentially declining) to the total stellar, stellar UV, TIR and PAH emission for the model (attenuation values fixed to those at the center of NGC~628). These plots show that the fractional PAH luminosity from the young stellar population is indeed higher than that of the total TIR, due to the stronger UV absorption of these small molecules. However, both have significant emission attributed to heating from stars older than 100~Myr as shown by the fractions shown for the lowest bar of each plot of Fig.~\ref{fig:fractional}, 26\% and 35\%. While these fractions of old-star heated PAH emission are smaller than what we derive observationally for NGC~628 (43\%), shifting to a reasonable twice-as-high escape fraction for the UV as for the ionizing photons lowers the observed fraction to 30\% (see Section~5.1), similar to the results of the model. The model and observations appear to be consistent within the uncertainties.

\section{Tests of systematics}

\subsection{UV escape fraction}
Arguments can be made as to why the UV escape fraction should be similar to that of the ionizing photons, but because different physical mechanisms are in play, the ratio between the fractions is unlikely to be unity, as we have assumed. In fact, the observed diffuse \fha fraction is only 28\% while the escape fraction calculated for the typical \hii-region $A_{\mathrm{V}}$ of 0.7 at 1500\AA\xspace is 49\% (applying the starburst attenuation law from \citealp{calzetti94}). It is thus likely that the UV escape fraction is larger than that for the ionizing photons. Assuming twice the rate of escape for UV photons compared to ionizing photons changes dust \feight fraction from 62\% to 53\% for $>$10Myr and from 43\% to 30\% for $>$100Myr stars. This change obviously lowers the \feight attributed to old stars, to a level more in line with our simple model. Our observations thus suggest that the old-star powered 8\micron dust emission fraction lies within these ranges. 

\subsection{Stellar subtraction at 8\micron}
A fraction ($\eta_{8\mu\mathrm{m}}$) of the 3.6 \micron IRAC band intensity, $I_{\mathrm{\nu}}(3.6\;\mathrm{\mu m})$, is subtracted from  the $I_{\mathrm{\nu}}(8\;\mathrm{\mu m})$ image in order to obtain a stellar-subtracted 8 \micron image. We test values of $\eta_{8\mu\mathrm{m}}$ from 0.22-0.30, centered on the adopted value of 0.255 of \citet{helou04}. Non-\hii region dust \feight fluxes are about 3\% greater in the $\eta_{8\mu\mathrm{m}}$=0.22 images than the $\eta_{8\mu\mathrm{m}}$=0.30 images, while the \hii region dust \feight  are only about 1\% greater. As both of these quantities simultaneously increase, the non-\hii region dust \feight fraction only ends up changing by ~0.5\%, a negligible amount.

The non-\hii and \hii region \feight$_\mathrm{dust}$/\halpha ratios change by the full 3\% and 1\%, respectively. But these few-percent changes are small compared to radial decline seen in both the non-\hii and \hii region ratios.

\subsection{\halpha stellar subtraction}
Similarly, a fraction of the wide-band image (mostly stellar continuum) is subtracted from the original narrow-band \halpha image (containing both line and continuum emission) in order to obtain an \halpha line-only image. Here, we test the effect of this stellar subtraction on our results. In Section~2.2, we noted that the mean ratio between the narrow-band and broad-band fluxes for 84 foreground stars was 0.806 with a standard deviation of 0.036. We use the $\pm1\sigma$ values to determine how different our results could be with different $\eta_{\mathrm{H}\alpha}$. 

The subtraction affects the non-\hii regions (-10\%, 15\%) more than the bright \hii regions (-2.5\%, 1\%) and especially in the galaxy center where the starlight is most concentrated. But based upon inspection, the different \halpha subtractions do not significantly affect the spatial extent or number of regions selected by \hiiphot. Overall, the non-\hii region \fha fraction changes from 34\% to 40\% for the under- and over-subtracted images (36\% for best-value) if no extinction correction is applied (to either \hii or diffuse regions), or only from 22\% to 27\% (24\% for best-value) if our preferred extinction correction is applied.  So, as with the 8\micron stellar subtraction experiment, it appears that the choice of \halpha stellar subtraction value does not strongly influence our results. 

\subsection{\halpha image properties}
Two properties of the \halpha image (resolution and noise) influence how \hiiphot identifies HII regions. The resolution here was set to match that of the 8\micron images, at 2.82$\arcsec$ = 98~pc at the 7.2~Mpc distance of NCC~628. Reducing the resolution to 4.23$\arcsec$(148~pc, in both the \halpha and 8\micron images) increases the diffuse \fha fraction from 24\% to 27\% and thus ends up decreasing the fraction of \feight heated by old stars from 47\% to 41\%. Testing the effect of using a finer scale would be more informative, but the fact that there is such a decrease as we degrade the resolution signals that even less of the \halpha emission may truly lie outside of \hii regions and thus even more of the 8\micron emission is in fact heated by old stars. Thus our conclusion that this number is a lower limit will hold.

We add Poisson noise in both the narrowband and broadband images separately and re-subtract to obtain a line-only image. We add more and more noise (mimicking a proportional decrease in exposure time between narrow and broad band images) until a rms noise level of 20 pc cm$^{-6}$ is obtained in the line-only image. Using such noisier data does not change any of our results within a 2\% margin. 

Increasing the noise by a significant factor was a less important effect, producing only a 2\% change in the final estimate of the fraction of \feight powered by old stars.

\subsection{Dust attenuation effects}

\begin{figure*}
\begin{center}
\includegraphics[width=15.5cm]{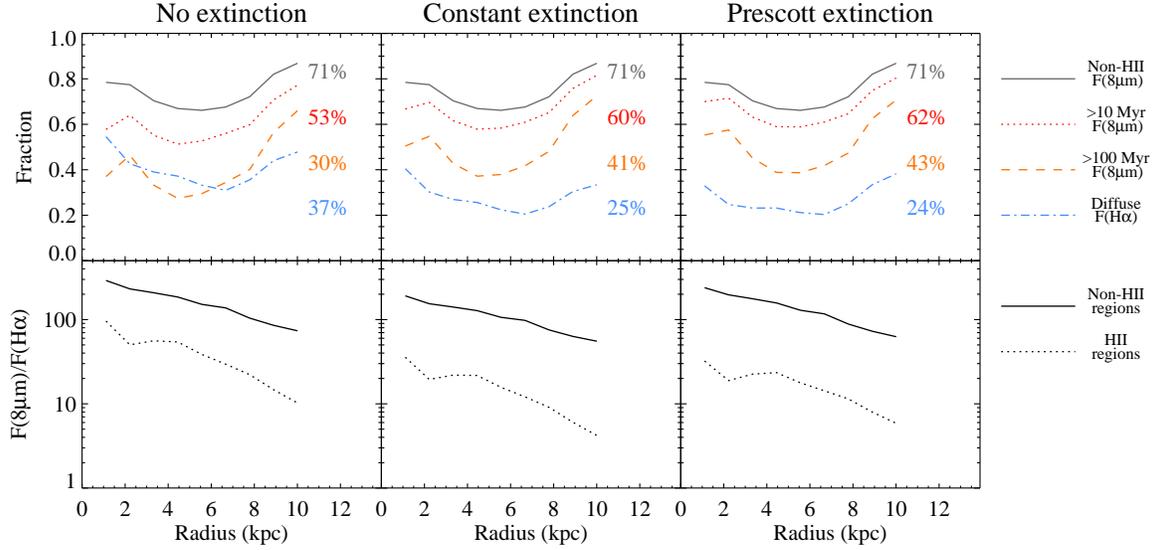}
\caption{Differences between different extinction corrections. The left column shows the results when no extinction correction is applied. The middle and right columns both apply a radially declining extinction correction to the observed non-\hii region H$\alpha$ emission, but the middle column applies a uniform extinction correction to the \hii regions from \citet{sanchez11} while the right column applies the radially declining extinction correction found in \citet{prescott07}. 
The colors and line styles are identical to Fig.~\ref{fig:results} and noted in the legend.}
\label{fig:extcorr}
\end{center}
\end{figure*}

Correcting the \halpha emission for dust attenuation significantly changes the non-\hii region (i.e. `diffuse') fraction of \fha. This effect can be seen clearly by following the blue line in the upper panels of Fig.~\ref{fig:extcorr}. With absolutely no extinction correction applied, the non-\hii \fha fraction is 37\%. With the \hii region correction from \citet{sanchez11} and the non-\hii region correction from the \citet{aniano12} dust map, the fraction is 25\%. With our adopted extinction correction of \citet{prescott07} (corrected via \citealt{calzetti07}) for \hii regions and again the dust map for the non-\hii regions, it is only 24\%. Thus, considering the differential extinction between within and outside of \hii regions is very important to determining the intrinsic amount of diffuse \halpha emission. Values may therefore be lower than the $\approx50$\% commonly reported by studies that do not take such differential extinction into account.

As the amount of diffuse \halpha emission determines how much diffuse 8\micron we attribute to emission escaping from \hii regions and thus associated to young stars, the extinction correction noticeably propagates through to our 8\micron fractions attributed to $>$10~Myr and $>$100~Myr populations, as seen in Fig.~\ref{fig:extcorr}. Accounting for differential extinction between the \hii regions and diffuse regions leaves about 10\% more 8\micron emission to be attributed to old stellar populations. Thus the extinction correction is an important systematic, although the difference between our two adopted extinction corrections is fairly minimal, only at the 2\% level.

\subsection{Emission measure cutoff effects}

\begin{figure}
\begin{center}
\includegraphics[width=7.5cm]{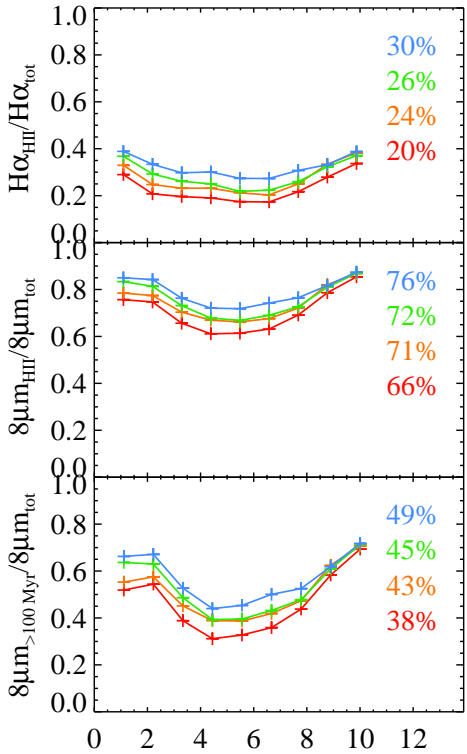}
\caption{Testing the effects of different emission measure cutoffs. The colors denote $\Delta$EM=1, 1.5, 2, 4 cm$^{-6}$, in rainbow order, from red to blue. The panels show from top to bottom: the non-\hii \halpha fraction, the non-\hii 8\micron dust fraction, and the 8\micron dust fraction heated by $>$100~Myr old stars. }
\label{fig:EM}
\end{center}
\end{figure}

The emission measure cutoff specifies the gradient in EM (measured in units of cm$^{-6}$ pc) at which the growth of \hii regions is terminated by \hiiphot. Lower EM cutoffs result in larger \hii regions. Figure~\ref{fig:EM} shows that decreasing the EM cutoff lowers both the \fha and \feight non-\hii fractions, as expected. These changes  are significant, with the non-\hii \fha fraction decreasing from 30\% to 20\% (Prescott extinction correction applied), the non-\hii \feight fraction from 76\% to 66\% and the lower limit on the \feight from old stars from 49\% to 38\%. However, inspection shows that the \hii regions for $\Delta$EM of 4 misses much of their contiguous \halpha emission, and those for $\Delta$EM of 1 are likely too large. Still, we expect the uncertainty on defining the exact boundary on \hii regions to contribute an $\approx$5\% uncertainty on our derived lower limit to the amount of 8\micron dust emission that is heated by stars older than 100~Myr.

\section{Conclusions}

Masking out \hii regions with the help of an \halpha image and the \hiiphot code, we find that 30-43\% of the non-stellar \feight in NGC~628 is due to PAH molecules and dust heated by stars older than 100~Myr. Our analysis shows that 38-47\% is heated by stars under 10~Myr, while 21-23\% is heated by stars between 10 and 100~Myr. The major parameter contributing to the ranges of these estimates is the unknown escape fraction of UV photons compared to ionizing photons. 

We thoroughly investigate the systematic effects of background subtraction, stellar emission subtraction (for both 8\micron and \halpha images), extinction correction and emission measure cutoff and estimate a combined systematic uncertainty of $\approx$10\% on the lower limit for non-star formation related \feight. Thus this work rigorously establishes that a significant fraction of \feight in NGC~628 is not due to heating from the youngest population of stars. The quantity of this non-star formation related emission may be different from galaxy to galaxy, complicating measures of the star formation rate determined using 8\micron emission at both high and low redshift.

We also find that the \feight/\fha ratio declines with galactocentric radius in \hii regions by a factor 5.4 and outside of \hii regions by a factor 4.0.  This decline is likely caused by the easier destruction or more difficult formation of PAHs in low metallicity regimes. However, PAH abundance may not be the full answer. Changes in PAH population, in terms of size distribution or ionization state, could also cause a reduction in 8\micron band emission. 

In the process of this work, we find that the fraction of non-\hii region (commonly called diffuse) \halpha emission is lower than usually reported, taking into account the  higher preferential extinction of \hii regions. With our adopted extinction model, we find an intrinsic non-\hii region \fha fraction of only 24\%. This may be compared to 37\% when differential extinction between HII regions and
diffuse emission is not taken into account. 

Independent of the \hii region based analysis, we compare the power spectra of the 8\micron dust and \halpha images and find that the 8\micron dust image has more power at lower spatial frequencies (larger scales) compared to the \halpha image. This distribution confirms the more diffuse nature of the 8\micron dust emission. The power-law slope of the 8\micron dust power spectrum matches well with that of the \hi emission in NGC~628, suggesting that the 8\micron emission is more strongly linked to the \hi gas than to star formation.

\bibliographystyle{apj}
\bibliography{master.bib}

\appendix{}
\section{A model of PAH excitation by young and old stellar populations}
\subsection{Stellar spectral synthesis}

Stellar spectra corresponding to specified age ranges are created using the spectral synthesis code of \citet{bruzual03} with Padova 1994 evolutionary tracks and a Chabrier IMF. We test metallicities of solar and two sub-solar values (z=0.02, 0.008 and 0.004) and a range of star formation histories. The first SFH is constant and the second is an exponential model with a timescale of 5~Gyr. The third is a linear model, based upon equation 9 of \citet{fraternali12} and the $\gamma$ they determined specifically for NGC~628. The final SFH is based upon the stellar population reconstruction from panchromatic photometry in \citet{zou11}. This SFH is shown in their fig. 5 and peaks at 2~Gyr ago. We normalize all the SFHs to have a current SFR of 0.7 \Msun yr$^{-1}$ \citep{kennicutt11}. This normalization results in total stellar masses (integrating over SFH and accounting for stellar mass loss) of 9.6, 10.1, 10.0 and 9.5 in log(\msun) for the constant, exponential, linear and 2~Gyr-peaked SFHs. The measured log(\msun) for NGC~628 is only 9.6 \citep{skibba11}. Thus, the constant and 2~Gyr-peaked SFHs are accurately tuned to NGC~628, while the linear and exponential models have about three times too much stellar mass. Still, they may represent broadly the shape of NGC~628's SFH (perhaps a period of elevated SFR at the present) so we keep them for the analysis.

\subsection{Dust absorption: non-ionizing radiation}

In order to estimate the amount of PAH emission due to old stellar populations, we create a simple model based on separating the absorption of stellar energy due to PAHs from that of the total population of dust grains. We first assume that the old stellar populations are mixed with the dust while the youngest stellar populations follow the starburst attenuation law, which is broadly equivalent to a foreground (clumpy) screen. We change from the foreground screen (with optical depth $\tau^{\mathrm{scr}}_{\lambda}$) to the mixed dust geometry (with optical depth $\tau^{\mathrm{mix}}_{\lambda}$) assuming a timescale for young clusters to emerge from their natal clouds of 10~Myr. The following equation describes the fraction of absorbed light as a function of the screen and mixed optical depths  and the age of the stellar population (t, in Myr):

\begin{equation}
f_{\mathrm abs}(\tau,t) = \left[\left(1-e^{-\tau^{\mathrm{scr}}_{\lambda}}\right)-\left(1-\frac{1-e^{-\tau^{\mathrm{mix}}_{\lambda}}}{\tau^{\mathrm{mix}}_{\lambda}}\right)\right]e^{-t/10}+\left(1-\frac{1-e^{-\tau^{\mathrm{mix}}_{\lambda}}}{\tau^{\mathrm{mix}}_{\lambda}}\right).
\end{equation}
The variation of optical depth with wavelength is computed by the ratio of the absorption cross-section of all dust (including PAHs),  $\kappa_{\mathrm{tot}}(\lambda)$, normalized to that at V-band as follows:
\begin{equation}
\tau(\lambda) = \frac{\kappa_{\mathrm {tot}}(\lambda)}{\kappa_{\mathrm {tot}}(V)}\tau_{\mathrm{V}}.
\end{equation}
We note that this approach to the attenuation of an integrated stellar population is similar to that of \citet{dacunha08}, with an important difference being that we use a dust-model based cross-section instead of a simple power-law effective absorption curve.

With Equations~A1 and A2 determining $f_{\mathrm{abs}}$ as a function of stellar population age and wavelength for given $\tau_{\mathrm{V}}^{\mathrm{scr}}$ and $\tau_{\mathrm{V}}^{\mathrm{mix}}$, we may then calculate the TIR luminosity for stellar populations with ages between $t_{1}$ and $t_{2}$ as:
\begin{equation}
L_{\mathrm{TIR}}=\int^{t_{2}}_{t_{1}}dt\int^{\lambda_{2}}_{\lambda_{1}}d\lambda\frac{dL_{\lambda}}{dt}f_{\mathrm{abs}}(t,\lambda).
\end{equation}
For the PAH luminosity, we apply the ratio between the PAH-only cross-section to the total dust cross-section:
\begin{equation}
L_{\mathrm{PAH}}=\int^{t_{2}}_{t_{1}}dt\int^{\lambda_{2}}_{\lambda_{1}}d\lambda\frac{dL_{\lambda}}{dt}\frac{\kappa_{\mathrm{PAH}}(\lambda)}{\kappa_{\mathrm{tot}}(\lambda)}f_{\mathrm{abs}}(t,\lambda).
\end{equation}

\begin{figure}
\begin{center}
\includegraphics[width=7.5cm]{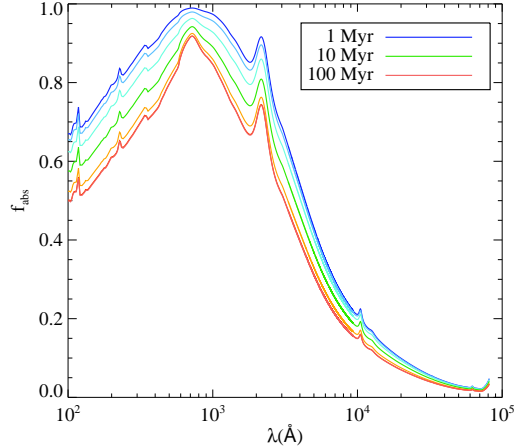}
\caption{{\bf Left:} The fraction of absorbed light versus wavelength for various age stellar populations, ranging from 1~Myr (blue) to 100~Myr (red). For these absorbed light fractions, we have assumed a \tauoldv=0.69 and a \tauyoungv=0.5, appropriate for the center of NGC~628. While these optical depths might seem backwards (higher optical depth towards older population), the different dust geometries assumed mean that the absorbed fraction is still higher towards the young population. }
\label{fig:fabs}
\end{center}
\end{figure}

\subsection{Dust absorption: ionizing radiation}

For ionizing photons, hydrogen atoms present a major source of opacity in addition to dust. We assume that \hii regions are radiation bounded and that a fraction of ionizing photons, $f_{\mathrm{ion}}$, ionize hydrogen. The remainder are absorbed by dust within the \hii region, which is assumed to be devoid of PAHs \citep[e.g.][]{giard94,churchwell06}. About 70\% of ionizations produce a Ly$\alpha$ photon upon recombination (valid for electron densities, $n_{\mathrm e} < 10^3$ cm$^{-3}$, \citealp{draine11}), which is then repeatedly scattered. After several scatterings, most of these Ly$\alpha$ photons will be absorbed by dust, while a small fraction, $f_{\mathrm {esc}}$, escape.
Let $Q_{0}$ be the rate of ionizing photons in s$^{-1}$ and $I_{\mathrm{H}}$ be the ionization energy of hydrogen ($2.179\times10^{-11}$ erg). If the average energy of an ionizing photon is approximately 1.2$I_{\mathrm{H}}$, then the TIR luminosity from within the \hii region from ionizing photons is:
\begin{equation}
L_{\mathrm{TIR,\;HII(ionizing)}}=Q_{0}I_{\mathrm{H}}\left[1.2\left(1-f_{\mathrm{ion}}\right) + 0.7*\frac{3}{4}f_{\mathrm{ion}}\left(1-f_{\mathrm{esc}}\right)\right].
\end{equation}
Note the factor $3/4$ gives the energy of the Ly$\alpha$ transition by the Rydberg formula.

\hii regions are surrounded by dusty PDRs. The line emission from any escaping Ly$\alpha$ as well as from optical hydrogen lines and metallic cooling lines should also be included here, as this is essentially reprocessed ionizing radiation. We assume that the PDR has a covering fraction ($\phi_{\mathrm{PDR}}$) of 0.7. For the optical recombination emission, we know that the 30\% of recombinations that do not produce a Ly$\alpha$ photon will produce a two-photon decay from the 2s state. Before decaying to the 2s state, the electron must have already lost 0.25$I_{\mathrm{H}}$ through recombination lines. For the metallic cooling lines, we assume they produce 0.1$I_{\mathrm{H}}$ per hydrogen ionization. Combining all of these terms, we have:
\begin{equation}
L_{\mathrm{TIR,\;PDR(ionizing)}}=Q_{0}I_{\mathrm{H}}f_{\mathrm{ion}}\phi_{\mathrm{PDR}}\left[f_{\mathrm{esc}} + 0.3*\frac{3}{4} + 0.25 + 0.1)\right].
\end{equation}
PAHs do exist within the PDR, so we also have:
\begin{equation}
L_{\mathrm{PAH,\;PDR(ionizing)}}=Q_{0}I_{\mathrm{H}}f_{\mathrm{ion}}\phi_{\mathrm{PDR}}\left[\frac{\kappa_{\mathrm{PAH}}(\mathrm{Ly\alpha})}{\kappa_{\mathrm{tot}}(\mathrm{Ly\alpha})}f_{\mathrm{esc}} +\frac{\kappa_{\mathrm{PAH}}(\mathrm{opt})}{\kappa_{\mathrm{tot}}(\mathrm{opt})}0.575\right].
\end{equation}

\begin{figure}
\begin{center}
\includegraphics[width=7.5cm]{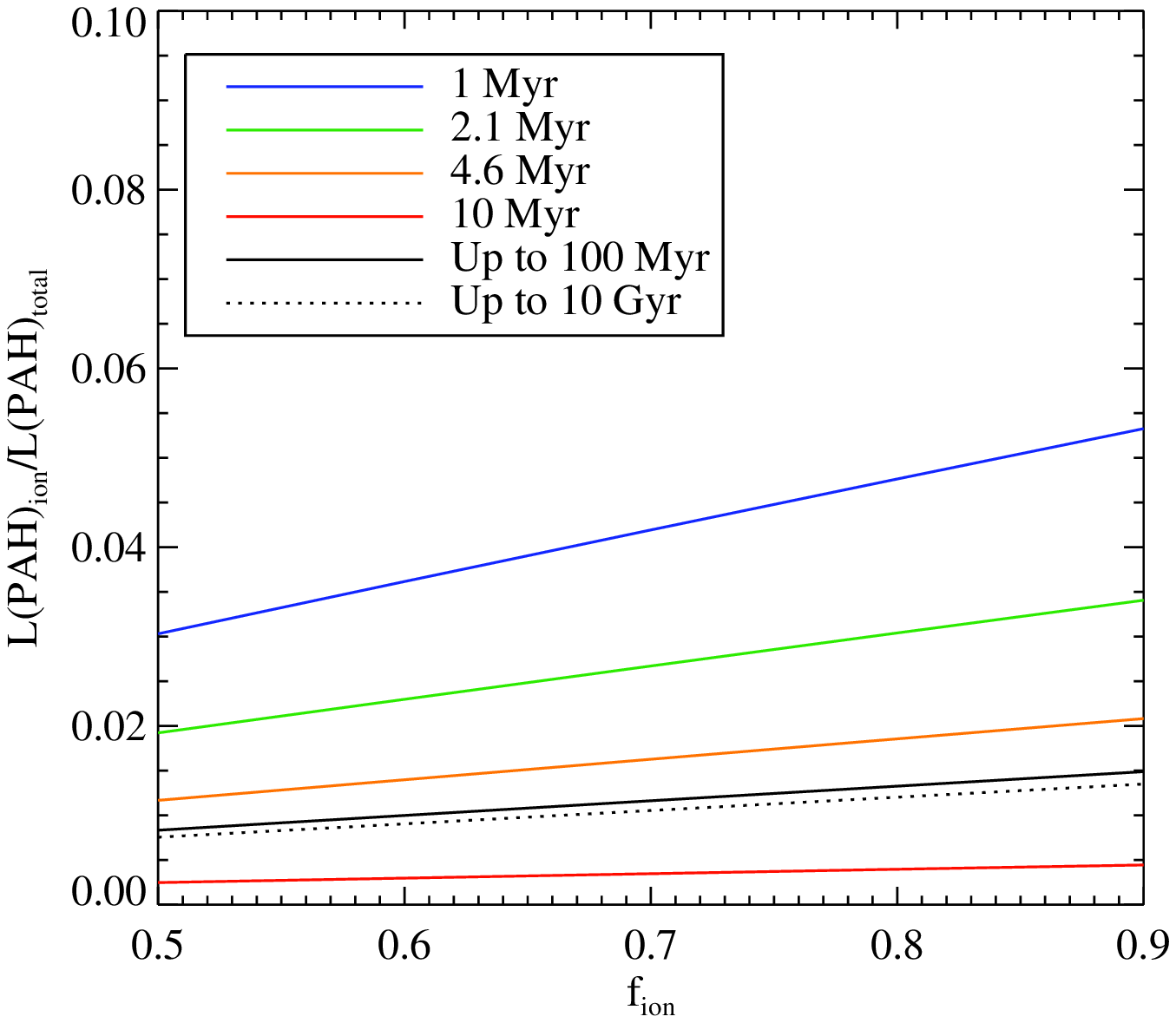}
\includegraphics[width=7.5cm]{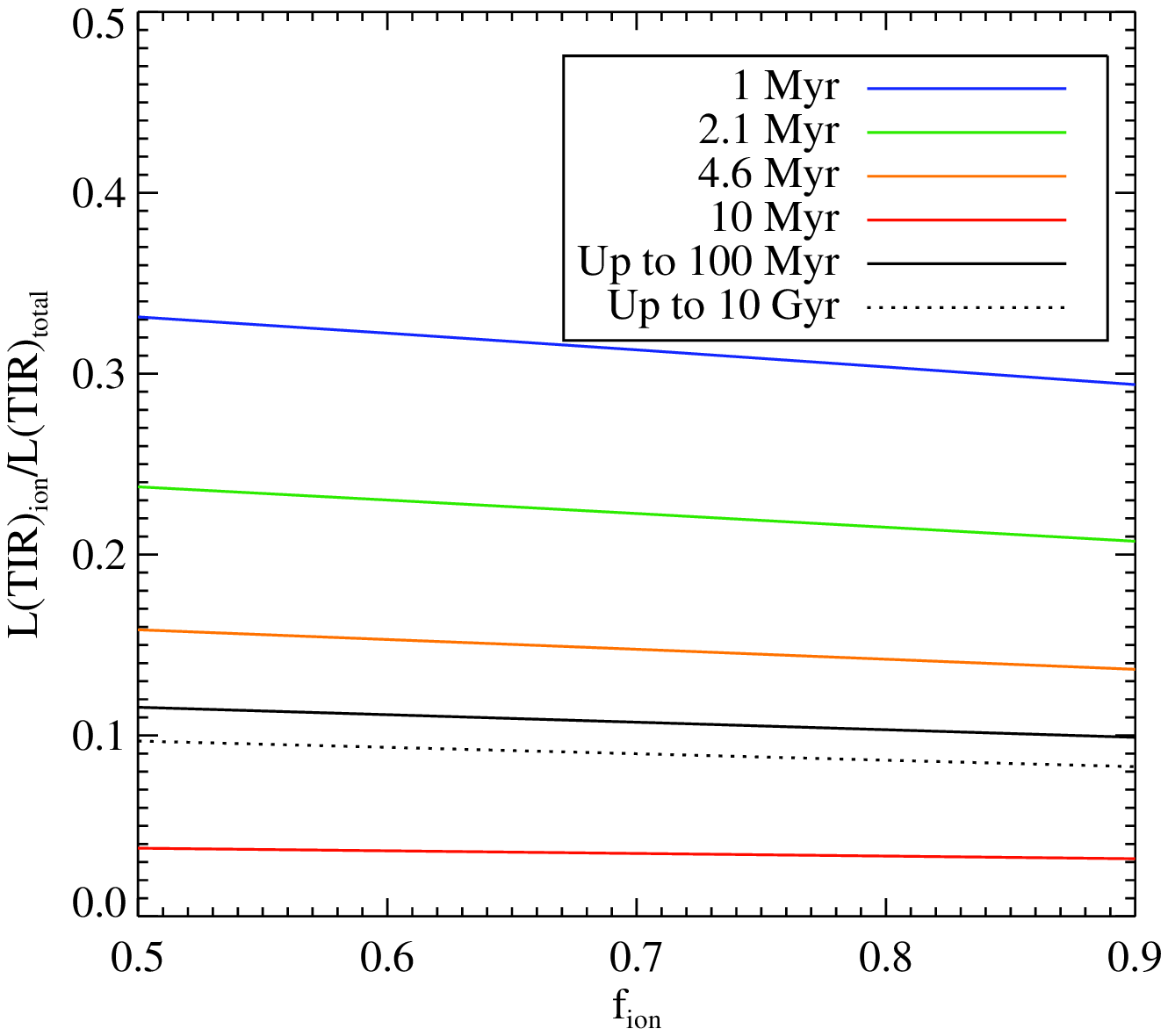}
\caption{The fraction of the total PAH (left) and TIR (right) emission contributed by ionizing photons as a function of the fraction of ionizing photons that actually ionize hydrogen. Different colored lines represent the fraction for different age stellar populations. The solid black line represents the fraction for a continuous SFR population up to 100~Myr while the dotted line does the same up to 10~Gyr.}
\label{fig:fabs}
\end{center}
\end{figure}

Because the dust within the \hii regions is assumed to be PAH-free, the ionizing radiation does not have a large contribution to the PAH emission. As shown in the left panel of Fig.~\ref{fig:fabs} (dotted black line), the escaping recombination and cooling radiation contributes less than 2\% to the total PAH emission (assuming a constant SFR 10~Gyr-old population), even with a high $f_{\mathrm{ion}}$ of 0.9. The TIR emission from intercepted ionizing photon light (either directly or through lines) is approximately 10\% as seen in the right panel of Fig.~\ref{fig:fabs}. We note that this fraction agrees with the 5-10\% found for FIR emission from \hii regions within the Milky Way \citep{sodroski97}.

\subsection{Model results}

Assuming a value of $f_{\mathrm{ion}}=0.7$ and combining the non-ionizing and ionizing contributions we obtain both the TIR and PAH luminosity for young stars we have:
\begin{equation}
L_{\mathrm{TIR,\;young}}=1.01Q_{0}I_{\mathrm{H}}+\int^{100\;\mathrm{Myr}}_{0}dt\int^{10,000\mathrm{\AA}}_{912\mathrm{\AA}}d\lambda\frac{dL_{\lambda}}{dt}f_{\mathrm{abs}}(t,\lambda)
\end{equation}
and
\begin{equation}
L_{\mathrm{PAH,\;young}}=0.057Q_{0}I_{\mathrm{H}}+\int^{100\;\mathrm{Myr}}_{0}dt\int^{10,000\mathrm{\AA}}_{912\mathrm{\AA}}d\lambda\frac{dL_{\lambda}}{dt}\frac{\kappa_{\mathrm{PAH}}(\lambda)}{\kappa_{\mathrm{TIR}}(\lambda)}f_{\mathrm{abs}}(t,\lambda).
\end{equation}
The old population luminosities may be calculated using Equations A3 and A4 with the wavelength range 912-10000\AA\xspace and the stellar population age range 100~Myr-10~Gyr. 

\begin{figure}
\begin{center}
\includegraphics[width=7.5cm]{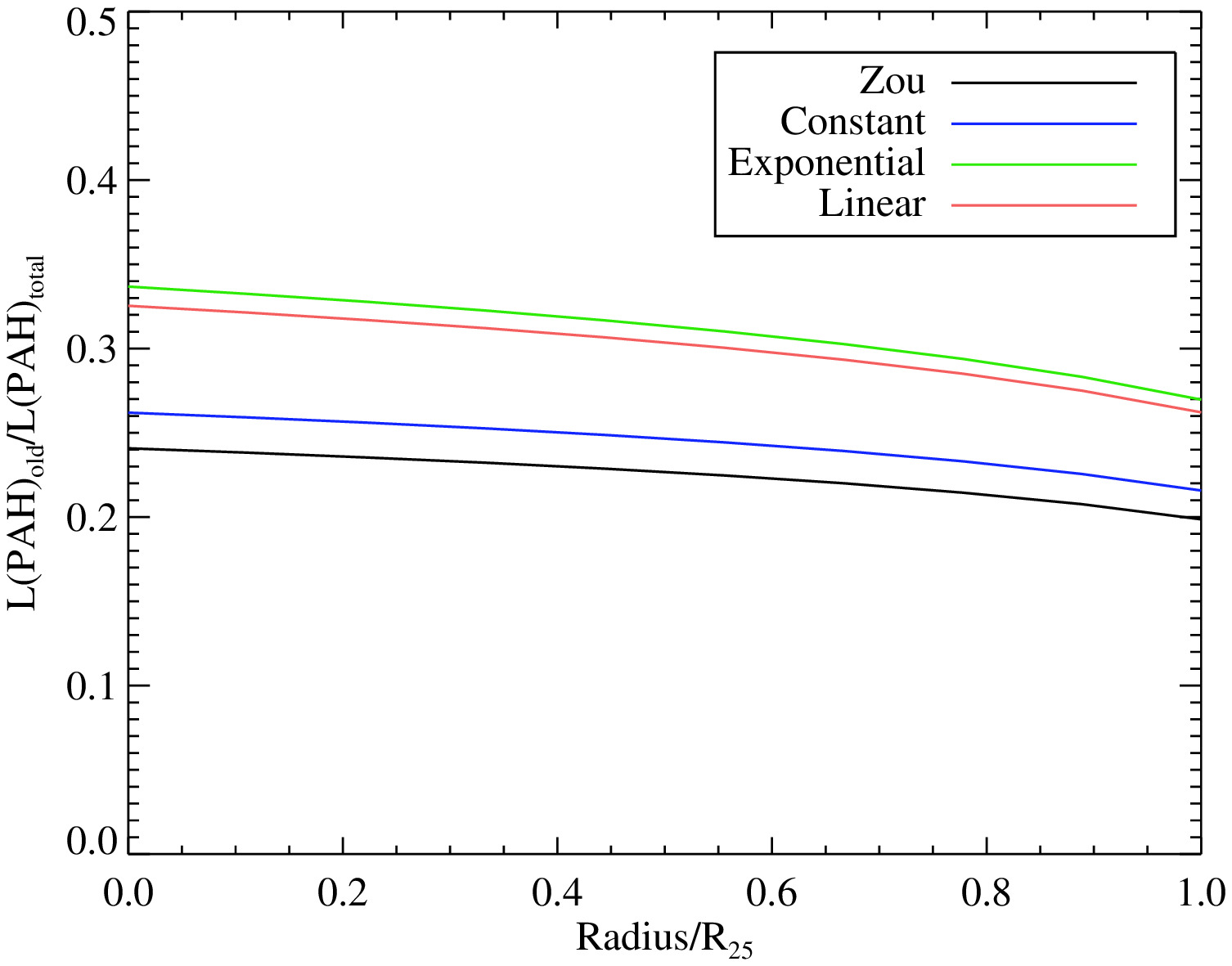}
\includegraphics[width=7.5cm]{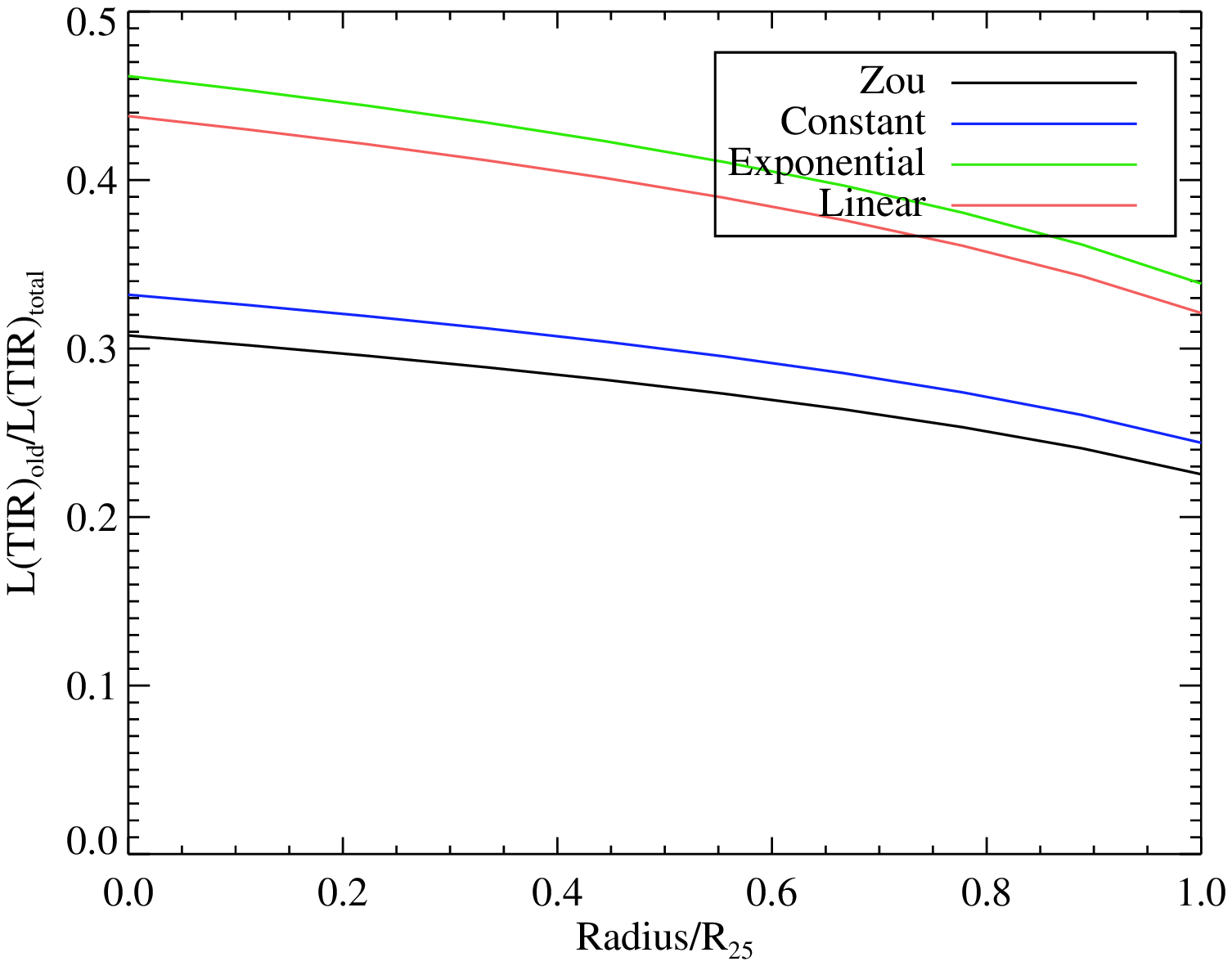}
\includegraphics[width=7.5cm]{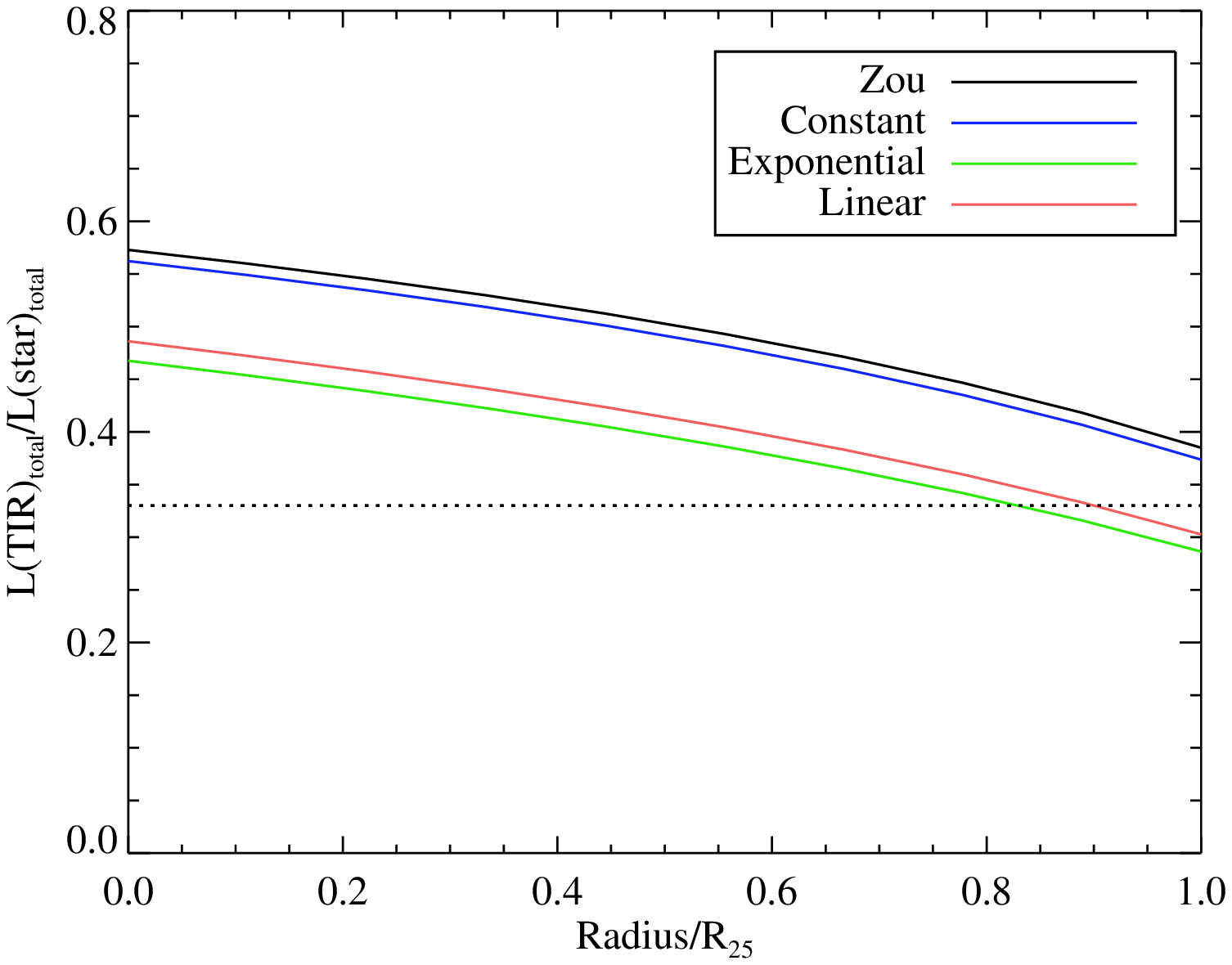}
\includegraphics[width=7.5cm]{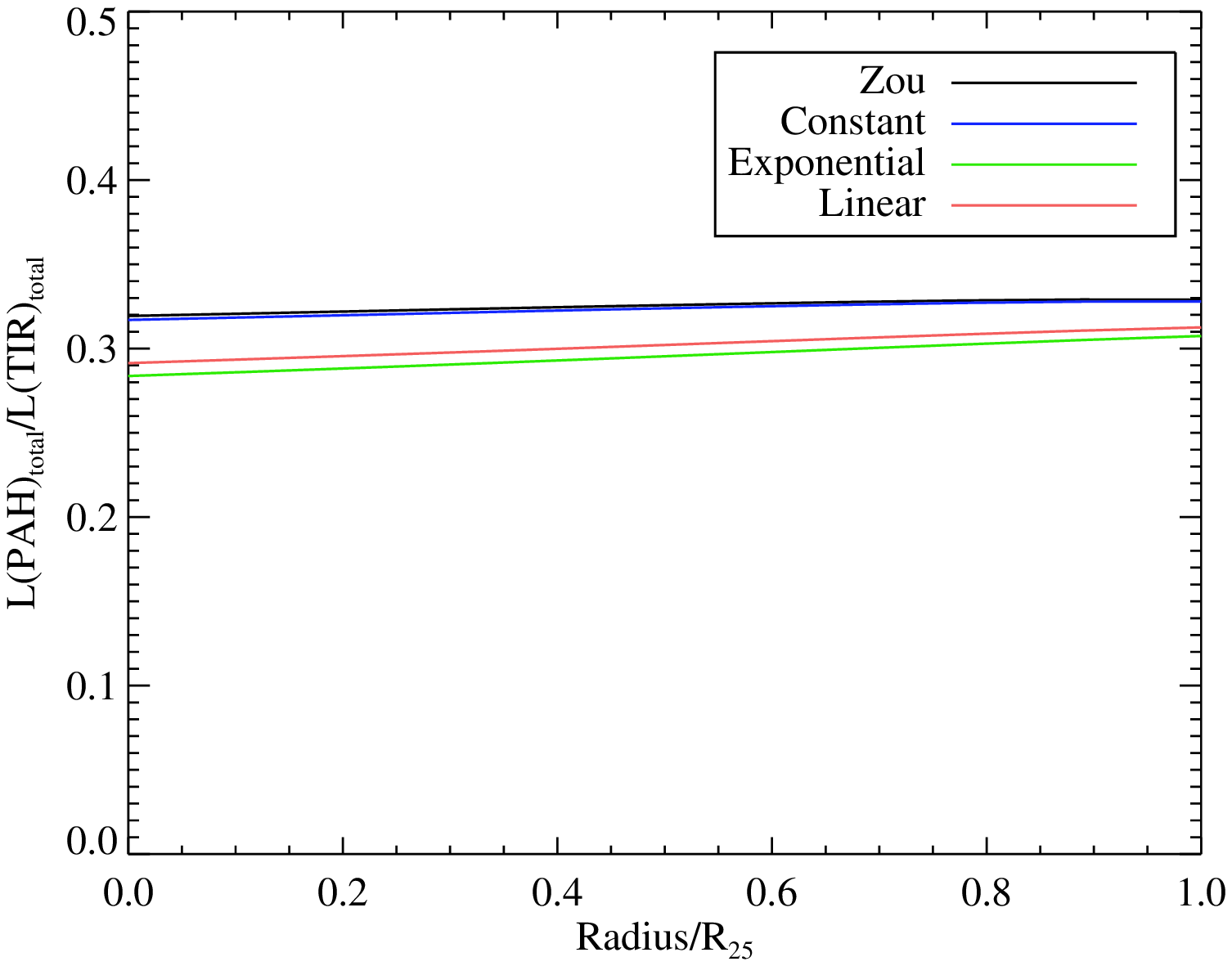}
\caption{{\bf Upper left:} The fraction of PAH luminosity contributed by old stars. {\bf Upper right:} The fraction of TIR luminosity contributed by old stars. {\bf Lower left:} The ratio of TIR to unattenuated stellar luminosity. The dotted line represents the global value for NGC~628 from \citet{skibba11}. {\bf Lower right:} The ratio of PAH to TIR luminosity. In all plots, the solid lines assume that the old population attenuation declines proportionally to that of the \hii regions while the dashed lines assume the old population's attenuation exponentially declines as per \citet{white00}. The different colors reflect the different SFHs as shown in the legend.}

\label{fig:withrad}
\end{center}
\end{figure}

The choices for \tauscr and \taumix used in Eqn.~A1 should be set to match NGC~628. From Section~3.2, we have $A$(\halpha)$=1.03-0.53 R/\mathrm{R}_{25}$ for the \hii regions. Assuming the continuum is less attenuated than the \halpha emission by the factor 0.44 \citep{calzetti94} and transforming to V-band and to optical depth instead of magnitudes, we have \tauscr$=0.51-0.26 R/\mathrm{R}_{25}$. For the older populations, we again adopt the prescription for the disk regions from Section~3.2 (linearly declining with radius), but this time use the full attenuation through the disk (and neglecting Milky Way foreground extinction). As the stellar photons exciting PAHs are traveling in all directions, the effective dust optical depth for absorption of these photons is likely to be comparable to the full-thickness optical depth, hence adopting that value. Using the dust surface densities from \citet{aniano12} and converting to optical depth, we parametrize \taumix$=1.20-0.92 R/\mathrm{R}_{25}$. (The different geometry assumed for young and old regions means that the higher \taumix values give similar or lower absorbed fractions of stellar radiation than the lower \tauscr.) These parameterizations of \taumix and \tauscr with radius result in Fig.~\ref{fig:withrad}, which shows the radial behavior of the fractions of PAH and TIR luminosity contributed by old stars (top left and right, respectively), the ratio of TIR to stellar luminosity (bottom left) and the ratio of PAH to TIR luminosity (bottom right). Fig.~11 also exhibits the effect of different star formation histories, with four different SFHs shown as different colored lines.  

Observationally, the TIR to stellar ratio is determined to be 0.33 in \citet{skibba11} for NGC~628. We plot this value as a horizontal dotted line in the lower left panel of Fig.~\ref{fig:withrad}. The linear and exponential SFHs are close to this value at all radii, while the constant and Zou SFHs both predict too much infrared per unit stellar luminosity. All SFHs and \tauold choices have a L(PAH) to L(TIR) ratio of approximately 0.3. This is larger than the typical observed value, but the way that PAH emission is usually measured does not include the continuum emission under the PAH features nor any continuum emission emitted at longer ($> 20$\micron) wavelengths. Furthermore, we did not use a dust cross-section perfectly tuned to NGC~628, the observationally-determined fractional PAH content ($q_{\mathrm{PAH}}$) is actually about 80\% lower than in the dust model we used, thus the PAH to TIR ratio should also decrease by this amount.

As for the PAH luminosity from old stars, the linear and exponential SFHs are quite close to the global value we derive, although they do not follow the specific shape from our Fig.~\ref{fig:results}. We take this as an indication that it is indeed possible for approximately 30\% of the PAH luminosity to be powered by old stars. The upper right hand panel of Fig.~\ref{fig:withrad} shows the corresponding plot for the TIR. While the cross-section of the bigger grains is weighted more towards longer wavelengths, the contribution from within the \hii regions partly makes up for this lower sensitivity to young stars and the fraction of TIR emission from old stars is only about 10\% more than for PAHs.

\acknowledgements{This work has been partially supported by the NASA ADP grant  NNX10AD08G and NSF grant AST 1008570. We thank the anonymous referee for his or her helpful comments which improved the paper. }

\label{lastpage}

\end{document}